\documentclass[twocolumn,aps,prc,superscriptaddress,showpacs,floatfix,longbibliography]{revtex4-1}
\usepackage{url}
\usepackage{cancel}
\usepackage[colorlinks,linkcolor=blue,citecolor=blue,filecolor=black,urlcolor=blue]{hyperref}
\usepackage{epsfig,graphics}
\usepackage{graphicx}
\usepackage{dcolumn}
\usepackage{bm}
\usepackage[usenames]{color}
\usepackage{amssymb}
\usepackage{amsmath}
\usepackage{multirow}
\usepackage{float}
\usepackage{harpoon}
\usepackage{MnSymbol}
\usepackage{appendix}
\usepackage{color}
\usepackage{hyperref}
\usepackage{cleveref}
\usepackage{ulem} 

\newcommand{\sqrtsnn}{\mbox{$\sqrt{s^{}_{\mathrm{NN}}}$}}
\newcommand{\pT} {p_{\mathrm{T}}}
\newcommand{\lr}[1]{\left\langle #1\right\rangle}
\newcommand{\pbpb}{$^{208}$Pb+$^{208}$Pb}
\newcommand{\auau}{$^{197}$Au+$^{197}$Au}
\newcommand{\ruru}{$^{96}$Ru+$^{96}$Ru}
\newcommand{\zrzr}{$^{96}$Zr+$^{96}$Zr}
\newcommand{\nch}{N_{\mathrm{ch}}}
\newcommand{\npart}{N_{\mathrm{part}}}
\newcommand{\Ns}{N_{\mathrm{s}}}
\newcommand{\pnbd}{p_{\mathrm{nbd}}}

\begin{document}

\title{Free spectator nucleons in ultracentral relativistic heavy-ion collisions as a probe of neutron skin}
\author{Lu-Meng Liu}
\affiliation{School of Physical Sciences, University of Chinese Academy of Sciences, Beijing 100049, China}
\author{Chun-Jian Zhang}
\affiliation{Department of Chemistry, Stony Brook University, Stony Brook, NY 11794, USA}
\author{Jun Xu}\email[Correspond to\ ]{xujun@zjlab.org.cn}
\affiliation{Shanghai Advanced Research Institute, Chinese Academy of Sciences, Shanghai 201210, China}
\affiliation{Shanghai Institute of Applied Physics, Chinese Academy of Sciences, Shanghai 201800, China}
\author{Jiangyong Jia}\email[Correspond to\ ]{jiangyong.jia@stonybrook.edu}
\affiliation{Department of Chemistry, Stony Brook University, Stony Brook, NY 11794, USA}
\affiliation{Physics Department, Brookhaven National Laboratory, Upton, NY 11976, USA}
\author{Guang-Xiong Peng}
\affiliation{School of Nuclear Science and Technology, University of Chinese Academy of Sciences, Beijing 100049, China}
\affiliation{Theoretical Physics Center for Science Facilities, Institute of High Energy Physics, Beijing 100049, China}
\affiliation{Synergetic Innovation Center for Quantum Effects $\&$ Applications, Hunan Normal University, Changsha 410081, China}
\date{\today}

\begin{abstract}
Besides the yield ratio of free spectator neutrons produced in ultracentral \zrzr\ to \ruru\ collisions~\cite{Liu:2022kvz}, we propose that the yield ratio $N_n/N_p$ of free spectator neutrons to protons in a single collision system at RHIC and LHC can be a more sensitive probe of the neutron-skin thickness $\Delta r_{\mathrm{np}}$ and the slope parameter $L$ of the symmetry energy. The idea is demonstrated based on the proton and neutron density distributions of colliding nuclei obtained from Skyrme-Hartree-Fock-Bogolyubov calculations, and a Glauber model that provides information of spectator matter. The final spectator particles are produced from direct emission, clusterization by a minimum spanning tree algorithm or a Wigner function approach, and deexcitation of heavy clusters by GEMINI. A larger $\Delta r_{\mathrm{np}}$ associated with a larger $L$ value increases the isospin asymmetry of spectator matter and thus leads to a larger $N_n/N_p$, especially in ultracentral collisions where the multiplicity of free nucleons are free from the uncertainties of cluster formation and deexcitation. We have further shown that the double ratio of $N_n/N_p$ in isobaric collision systems or in collisions by isotopes helps to cancel the detecting efficiency for protons. Effects from nuclear deformation and electromagnetic excitation are studied, and they are found to be subdominant compared to the expected sensitivity to $\Delta r_{\mathrm{np}}$.
\end{abstract}
\maketitle

\section{Introduction}
\label{sec:intro}

Understanding the nuclear matter equation of state (EOS) is one of the main goals in nuclear physics. Despite the efforts in the past decades, considerable uncertainties still remain in the isospin-dependent part of the EOS, i.e., the nuclear symmetry energy $E_{\mathrm{sym}}$ characterizing the energy difference between neutron-rich and neutron-poor systems. The accurate knowledge of $E_{\mathrm{sym}}$, particularly its density dependence, is important in understanding properties of neutron-rich nuclei, dynamics of nuclear reactions induced by radioactive beams, and phenomena in nuclear astrophysics~\cite{Steiner:2004fi,Lattimer:2006xb,Li:2008gp}, such as gravitational waves from compact star mergers. Among various probes of the nuclear symmetry energy, the neutron-skin thickness $\Delta r_{\mathrm{np}}$, defined as the difference between the neutron and proton root-mean-square (RMS) radii, is one of the most robust probes in constraining the slope parameter $L=3\rho_0 (dE_{\mathrm{sym}}/d\rho)_{\rho_0}$ of the symmetry energy around the saturation density $\rho_0$~\cite{Horowitz:2000xj,Furnstahl:2001un,Todd-Rutel:2005yzo,Centelles:2008vu,Zhang:2013wna,Xu:2020fdc}. In the past decades, the $\Delta r_{\mathrm{np}}$ has been measured experimentally through proton~\cite{Zenihiro:2010zz,Terashima:2008rb} and pion~\cite{Friedman:2012pa} scatterings, charge exchange reactions~\cite{Krasznahorkay:1999zz}, coherent pion photoproductions~\cite{Tarbert:2013jze}, and antiproton annihilations~\cite{Klos:2007is,Brown:2007zzc,Trzcinska:2001sy}, etc. More recently, experimental measurement of the $\Delta r_{\mathrm{np}}$ in $^{208}$Pb by parity-violating electron-nucleus scatterings favors a large value of $L$~\cite{Reed:2021nqk}. Although such measurement has less model dependence, there are still debates on the experimental method~\cite{PhysRevC.105.055503}, and the resulting large $L$ value differs from results based on other isovector observables, e.g., the electric dipole polarizability~\cite{Piekarewicz:2021jte}.

Heavy-ion collisions, of which the nucleon distribution determines the initial condition, are a useful way to measure the neutron-skin thickness as well. In low-energy heavy-ion collisions where the dynamics is mostly dominated by nucleon degree of freedom, the neutron/proton or triton/$^3$He yield ratio~\cite{Dai:2014rja} and the average isospin asymmetry of fragments~\cite{Dai:2015dua} were proposed to be probes of the neutron-skin thickness of colliding nuclei. However, such probes suffer from model dependence of transport simulations~\cite{Xu:2019hqg} and uncertainties of clusterization and fragmentation. In relativistic heavy-ion collisions, observables at midrapidities are also proposed as probes of the collective structure of colliding nuclei~\cite{Filip:2009zz,Shou:2014eya,Giacalone:2019pca,Jia:2021tzt}, including their neutron skins~\cite{Li:2019kkh}. For example, $^{96}$Ru+$^{96}$Ru and $^{96}$Zr+$^{96}$Zr collisions at $\sqrtsnn=200$ GeV were originally proposed to measure the chiral magnetic effect~\cite{STAR:2021mii}, but by product such collisions can also be used to measure the neutron-skin thickness of colliding nuclei through the ratio of midrapidity observables in the isobaric systems, such as that of the multiplicity distribution and/or $v_2$~\cite{Li:2019kkh,Jia:2021oyt}, the mean transverse momentum $\pT$~\cite{Xu:2021uar} and its fluctuations~\cite{PhysRevC.105.044905}, and $v_n$--$\pT$ correlations~\cite{Giacalone:2019pca,Bally:2021qys,Jia:2021wbq}. Since the final observables in relativistic heavy-ion collisions are affected by the complicated dynamics such as the relativistic hydrodynamics of the quark-gluon plasma, the hadronization, and the hadronic evolution, etc., these observables at midrapidities are also impacted by uncertainties in the theoretical description of the above processes as well.

In the present paper, we propose that the free spectator nucleons in ultracentral relativistic heavy-ion collisions can be used as a clean probe of the $\Delta r_{\mathrm{np}}$ of colliding nuclei. The free spectator neutrons can be measured by the zero-degree calorimeters (ZDC), and its distribution is mostly used to estimate the event centrality or the reaction plane~\cite{PHENIX:2000owy,ALICE:2013hur}. The free spectator protons can also be measured by instrumenting the forward region with dedicated detectors~\cite{Tarafdar:2014oua}. The advantage of choosing spectator nucleons in relativistic heavy-ion collisions as a probe is that the spectator matter passes by at the speed of light so that there are almost no interaction between the spectator matter and the participant matter, different from the situation in low-energy heavy-ion collisions. These free spectator nucleons are from the direct production of initial collisions and the deexcitation of heavy fragments. Generally, one expects that there are uncertainties from the calculation of the excitation energy of the fragments and also the deexcitation process, while in ultracentral heavy-ion collisions we will show that the results are insensitive to such uncertainties. Since free spectator neutrons in ultracentral heavy-ion collisions are mostly produced from the surface of the colliding nuclei, their multiplicity is causally related to the neutron-skin thickness, and a larger neutron skin intuitively leads to more free spectator neutrons than protons. As already observed in Ref.~\cite{Liu:2022kvz}, in \zrzr\ and \ruru\ collisions with the same mass but different isospin asymmetries, taking the ratio of the free spectator neutron yield in this isobaric system may largely cancel the uncertainties of theoretical modeling. In the present study, we will further show that the free spectator neutron/proton yield ratio in \auau\ and \pbpb\ collision systems at both RHIC and LHC energies can be a useful probe of the $\Delta r_{\mathrm{np}}$ of colliding nuclei if both free spectator neutrons and protons can be accurately measured, providing a complementary measurement of the $\Delta r_{\mathrm{np}}$ in heavy nuclei compared to other experiments, e.g., that for $^{208}$Pb by PREXII. Since the efficiency of detecting protons, whose motion is affected by the beam optics in the accelarator, is much lower than that of detecting neutrons, we further propose that the double ratio of free spectator neutron/proton yield ratio in \zrzr\ to \ruru\ collision system as an alternative probe of $\Delta r_{\mathrm{np}}$ and $L$, where the uncertainties of proton detecting efficiency are effectively cancelled out.

The rest part of the paper is organized as follows. Section~\ref{sec:theory} introduces the theoretical framework of the present study. We then present and discuss the numerical results of \zrzr, \ruru, \auau, and \pbpb\ collision systems at both RHIC and LHC energies in Sec.~\ref{sec:results}. We conclude and outlook in Sec.~\ref{sec:summary}.

\section{Theoretical framework}
\label{sec:theory}

In the present study, we first employ the Skyrme-Hartree-Fock (SHF) model to obtain the nucleon distributions of colliding nuclei. Based on such initial spatial distributions, the Glauber model is used to fit the charged particle multiplicity and to determine the spectator nucleons. We then apply a minimum spanning tree algorithm to determine heavy fragments and the direct production of free nucleons, and a Wigner function approach to describe the production of light clusters from the coalescence of free nucleons. With the excitation energies of clusters calculated through the test-particle method based on a simplified SHF energy-density functional, the GEMINI model is then used to deexcite heavy fragments. We also discuss the possible effect of the electromagnetic field on the excitation energy of heavy fragments in spectator matter.

\subsection{Skyrme-Hartree-Fock model}
\label{sec:SHF}

In the standard SHF model, the effective interaction between two nucleons at the positions $\vec{r}_1$ and $\vec{r}_2$ is expressed as
\begin{eqnarray}\label{SHFI}
v(\vec{r}_1,\vec{r}_2) &=& t_0(1+x_0P_\sigma)\delta(\vec{r}) \notag \\
&+& \frac{1}{2} t_1(1+x_1P_\sigma)[{\vec{k}'^2}\delta(\vec{r})+\delta(\vec{r})\vec{k}^2] \notag\\
&+&t_2(1+x_2P_\sigma)\vec{k}' \cdot \delta(\vec{r})\vec{k} \notag\\
&+&\frac{1}{6}t_3(1+x_3P_\sigma)\rho^\alpha(\vec{R})\delta(\vec{r}) \notag\\
&+& i W_0(\vec{\sigma}_1+\vec{\sigma_2})[\vec{k}' \times \delta(\vec{r})\vec{k}].
\end{eqnarray}
In the above, $\vec{r}=\vec{r}_1-\vec{r}_2$ is the relative coordinate of the two nucleons,  $\vec{R}=(\vec{r}_1+\vec{r}_2)/2$ is their central coordinate with $\rho(\vec{R})$ being the nucleon density there, $\vec{k}=(\nabla_1-\nabla_2)/2i$ is the relative momentum operator and $\vec{k}'$ is its complex conjugate acting on the left, and $P_\sigma=(1+\vec{\sigma}_1 \cdot \vec{\sigma}_2)/2$ is the spin exchange operator, with $\vec{\sigma}_{1(2)}$ being the Pauli matrices acting on nucleon 1(2). The 9 parameters in the Skyrme interaction $t_0$, $t_1$, $t_2$, $t_3$, $x_0$, $x_1$, $x_2$, $x_3$, and $\alpha$ can be expressed analytically in terms of 9 macroscopic quantities including the slope parameter $L$ of the symmetry energy. The model allows us to vary $L$ while keeping the other macroscopic quantities fixed at their empirical values~\cite{Chen:2010qx} that give a reasonable description of global nuclear structure data. Based on the Skyrme interaction, the energy-density functional can then be obtained using the Hartree-Fock method, and the single-particle Hamiltonian is obtained using the variational principle, with the Coulomb interaction also explicitly included. Solving the Schr\"odinger equation gives the wave functions of constituent neutrons and protons and thus their density distributions $\rho_\tau(\vec{r})$ with $\tau=n,p$ being the isospin index. As an essential ingredient for the description of open shell nuclei, the paring correlations have also been incorporated in above standard procedure, and this is called the Skyrme-Hartree-Fock-Bogolyubov (SHFB) method. In the present study, we have used the open source SHFB code in Ref.~\cite{Stoitsov:2012ri}, where the effects of the axial symmetry deformation $\beta_{\lambda}$ are included by using the cylindrical transformed deformed harmonic oscillator basis.

For a given spatial distribution of nucleons in a nucleus, the neutron-skin thickness is defined as
\begin{equation}
    \Delta r_{\mathrm{np}} = \sqrt{\lr{r_\mathrm{n}^2}}-\sqrt{\lr{r_\mathrm{p}^2}},
\end{equation}
where
\begin{equation}
    \sqrt{\lr{r_\mathrm{\tau}^2}} = \left(\frac{\int \rho_{\tau}(\vec{r}) r^2 d^3{r}}{\int \rho_{\tau}(\vec{r}) d^3{r} }\right)^{1/2}
\end{equation}
is the RMS radius for nucleons with isospin index $\tau$.

The intrinsic multipole moments for nucleons with isospin index $\tau$ in the case of axial symmetry can be calculated from
\begin{equation}
Q_{\lambda, \tau} =  \int \rho_{\tau}(\vec{r}) r^\lambda Y_{\lambda 0}(\theta) d^3{r},
\end{equation}
where $Y_{\lambda 0}(\theta)$ is reduced from the standard spherical harmonics $Y_{\lambda \mu}(\theta,\phi)$ with magnetic quantum number $\mu=0$. The corresponding deformation parameter $\beta_{\lambda,\tau}$ is then given by~\cite{Gambhir:1990uyn,Wang:2021tjg}
\begin{equation}
    \beta_{\lambda, \tau} =  \frac{4 \pi Q_{\lambda, \tau}}{3 N_\tau R^\lambda},
\end{equation}
where $R = r_0 A^{1/3}$ with $r_0 = 1.2$ fm is the empirical radius of a nucleus with total nucleon number $A$, and $N_\tau=\int \rho_{\tau}(\vec{r}) d^3{r}$ is the total number of nucleons with isospin index $\tau$. The above definition is for the deformation of neutron or proton distributions, and can be easily generalized to that of total nucleons. In the deformed SHFB calculation~\cite{Stoitsov:2012ri}, we can get the constrained $Q_{\lambda}$ of the resulting density distribution by the linear constraint method based on the approximation of the Random Phase Approximation matrix. An analysis of the ratio of $v_2$ and $v_3$ in the isobaric collisions favors $\beta_2=0.16$ for $^{96}$Ru and $\beta_2=0.06$ and $\beta_3=0.2$ for $^{96}$Zr~\cite{Zhang:2021kxj}. For the quadrupole deformation parameter of $^{197}$Au, we choose it to be $\beta_{\mathrm{2Au}}=-0.15$ favored by the $v_2$ data at RHIC energy from colliding nuclei with different $\beta_2$~\cite{Giacalone:2021udy}, and also consistent with theoretical calculations based on the finite-range droplet macroscopic model~\cite{Moller:2015fba}. For the double magic nucleus $^{208}$Pb, it is a spherical one.

\subsection{Glauber model}
\label{sec:Glauber}

The nucleus-nucleus collisions are simulated using a Monte-Carlo Glauber model~\cite{Miller:2007ri}, which is well valid in relativistic heavy-ion collisions. In each collision event, the colliding nuclei are placed with a random impact parameter, and their orientations with respect to the beam direction are uniformly randomized. The spatial coordinates of neutrons and protons are sampled according to the density distributions obtained from the SHFB calculation mentioned above, while their momenta are sampled in the isospin-dependent Fermi sphere according to the local density of neutrons or protons. The momenta of spectator nucleons are kept unchanged after the inelastic collisions for participant nucleons. The nucleon-nucleon (NN) inelastic cross sections used in this study at different collision energies are listed in Table~\ref{tab:crosssection}.
From the above information, the participant nucleons and spectator nucleons are identified, and the quantities associated with participant matter, such as the numbers of participating nucleons $N_{\mathrm{part}}$ and NN binary collisions $N_{\mathrm{coll}}$, are calculated for each event. With the distribution of $N_{\mathrm{part}}$ and $N_{\mathrm{coll}}$, the measured distribution of charged-particle multiplicity $\nch$ according to the experimental procedure is fitted by using the two-component Glauber model~\cite{Kharzeev:2000ph}. In this way, the quantities associated with spectator matter can then be correlated with $\nch$ similar to the experiment.

\begin{table}[h!]
\centering
\caption{Values of the NN inelastic cross section $\sigma^{}_{\mathrm{NN}}$ used in the Glauber model at various \sqrtsnn.  }
\label{tab:crosssection}
\renewcommand\arraystretch{1.5}
\setlength{\tabcolsep}{5mm}
\begin{tabular}{cccc}
\hline\hline
$\sqrtsnn$ & 130 GeV & 200 GeV & 5.02 TeV\\
\hline
$\sigma^{}_{\mathrm{NN}}$ & 40 mb & 42 mb & 68 mb \\
\hline\hline
\end{tabular}
\end{table}

In the two-component Glauber model~\cite{Kharzeev:2000ph}, the number of sources that produce charged particles in each nucleus-nucleus collision event can be expressed as
\begin{eqnarray}
\label{N_s}
\Ns = (1-x)\frac{N_{\mathrm{part}}}{2}+xN_{\mathrm{coll}},
\end{eqnarray}
where the parameter $x$ represents the relative contribution to the charged-particle multiplicity $\nch$ from hard processes. The distribution of sources in the collision zone is described by the standard Glauber model for various collision systems through the distribution of $N_{\mathrm{part}}$ and $N_{\mathrm{coll}}$. The particle production $n$ from each source is assumed to follow a negative binomial distribution (NBD)
\begin{eqnarray}
\label{pnbd}
\pnbd(n;m,p) = \frac{(n+m-1)!}{(m-1)!n!} p^n(1-p)^m,~ p = \frac{\bar{n}}{\bar{n}+m},
\end{eqnarray}
where the parameter $\bar{n}$ ($m$) represents the average number of particle in (not in) the acceptance, so $p$ is the probability of a particle falling into the acceptance. The distribution of total charged-particle multiplicity is obtained as a superposition of the NBD from all sources~\cite{Zhou:2018fxx}, i.e.,
\begin{eqnarray}
\label{pnbdtotal}
p(\nch;\Ns) &=&\pnbd(n_1;m,p)\otimes \pnbd(n_2;m,p)\nonumber\\
&&\otimes \ldots \otimes \pnbd(n_{_{\Ns}};m,p)  \nonumber\\
&=& \pnbd(\nch; m\Ns,p), ~ \nch \equiv \sum_{i=1}^{\Ns} n_i,
\end{eqnarray}
where the additive nature of the NBD for convolution is used. Values of the parameters $x$, $\bar{n}$, and $m$ are adjusted to fit the $\nch$ distribution.

\subsection{Clusterization algorithms}
\label{sec:winger}

The spectator matter obtained from the Glauber model are further grouped into charged clusters and free nucleons based on a minimum spanning tree algorithm with respect to both coordinate and momentum space. Nucleons close in phase space, i.e., with the distance $\Delta r<\Delta r_{\mathrm{max}}$ and relative momentum $\Delta p<\Delta p_{\mathrm{max}}$, are assigned to the same cluster. Here $\Delta r_{\mathrm{max}}$ and $\Delta p_{\mathrm{max}}$ represent the scale of maximum distance and relative momentum between neighboring nucleons in the clusters, respectively.
In the present study, we set $\Delta r_{\mathrm{max}}=3$~fm and $\Delta p_{\mathrm{max}}=300$ MeV/$c$ as in Ref.~\cite{Li:1997rc}, which have been shown to give the best description of the experimental data of free spectator neutrons in ultracentral \auau\ collisions at $\sqrtsnn=130$ GeV~\cite{Liu:2022kvz}.

The above method is used to identify heavy clusters with $A \geq 4$. For spectator nucleons that do not form heavy clusters ($A \geq 4$), they still have chance to coalesce into light clusters with $A \leq 3$, i.e., deuterons, tritons, and $^3$He, and the formation probabilities are calculated according to the following Wigner functions~\cite{Chen:2003ava,Sun:2017ooe} in their center-of-mass (C.M.) frame, i.e.,
\begin{eqnarray}
f_d &=& 8g_d \exp{\left(-\frac{\rho^2}{\sigma_d^2}-p_\rho^2\sigma_d^2\right)}\label{wig1}, \\
f_{t/^3He} &=& 8^2 g_{t/^3\mathrm{He}} \exp{\left[-\left(\frac{\rho^2+\lambda^2}{\sigma_{t/^3\mathrm{He}}^2}\right)- (p_\rho^2+p_\lambda^2)\sigma_{t/^3\mathrm{He}}^2\right]}\label{wig2}, \nonumber\\
\end{eqnarray}
with $\vec{\rho}=(\vec{r}_1-\vec{r}_2)/\sqrt{2}$, $\vec{p}_\rho=(\vec{p}_1-\vec{p}_2)/\sqrt{2}$, $\vec{\lambda}=(\vec{r}_1+\vec{r}_2-2\vec{r}_3)/\sqrt{6}$, and $\vec{p}_\lambda=(\vec{p}_1+\vec{p}_2-2\vec{p}_3)/\sqrt{6}$ being the relative coordinates and momenta. $g_d=3/4$ ($g_{t/^3\mathrm{He}}=1/4$) is the statistical factor for spin 1/2 proton and neutron to form a spin 1 deuteron (spin 1/2 triton/$^3$He). The width of the Wigner function is chosen to be $\sigma_d=2.26$ fm, $\sigma_t=1.59$ fm, and $\sigma_{^3\mathrm{He}}=1.76$ fm, according to the radius of deuterons, tritons, and $^3$He~\cite{Ropke:2008qk}, respectively. This approach has been shown to describe well the production of light clusters in both low- ~\cite{Chen:2003ava} and high-energy~\cite{Zhao:2020irc} heavy-ion collisions.

In the calculation, we consider all possible combinations of neutrons and protons that can form light clusters. The numbers of formed light clusters are counted according to Eqs.~(\ref{wig1}) and (\ref{wig2}), and the weights of neutrons and protons are correspondingly subtracted. The total free spectator nucleons are composed of the remaining neutrons and protons that have not coalesced into light clusters, and those from the deexcitation of heavy clusters. The total light clusters are composed of the ones that are coalesced from directed produced neutrons and protons described above, and also those from the deexcitation of heavy clusters. The deexcitation of heavy clusters is described by the GEMINI model as discussed in the following subsection.

\subsection{GEMINI model}
\label{sec:gemini}

GEMINI is a Monte Carlo model to calculate the deexcitation of nuclear fragments, which allows all possible binary decay modes including light-particle evaporation and symmetric/asymmetric fission~\cite{Charity:1988zz,Charity:2010wk}. With the information of a given primary fragment including its proton number $Z$, mass number $A$, excitation energy $E^*$, and spin $J_\mathrm{CN}$, GEMINI deexcites the fragment by a series of sequential binary decays until such decays are impossible due to the energy conservation or improbable due to gamma-ray competition. The gamma-ray emission is also included in GEMINI to correctly model the termination of particle decay, since the partial decay widths for particle and gamma decay can be comparable at very low excitation energies. In the past few decades, GEMINI has been much updated, and it is widely used in the calculation of compound-nucleus (CN) decay. In dynamical simulations of heavy-ion collisions, GEMINI has been used to deexcite fragments based on the final phase-space information (see, e.g., Refs.~\cite{Wu:2019wfx,Dai:2015dua}), so that the experimental data relevant for nuclear clusters can be successfully reproduced.

In GEMINI, the evaporation of light particles is treated with the Hauser-Feshbach formalism~\cite{Hauser:1952zz} which explicitly takes into account the spin degree of freedom. The partial decay width of a compound nucleus of excitation energy $E^{*}$ and spin $J_\mathrm{CN}$ for the evaporation of particle $i$ is expressed as
\begin{multline}
\Gamma _{i}(E^*,J_\mathrm{CN})=\frac{1}{2\pi \rho _\mathrm{CN}\left( E^{* },J_\mathrm{CN}\right) }\int
d\varepsilon \sum_{J_{d}=0}^{\infty }
\\
\sum_{J=\left\vert
J_\mathrm{CN}-J_{d}\right\vert }^{J_\mathrm{CN}+J_{d}}\sum_{\ell=\left\vert
J-S_{i}\right\vert }^{J+S_{i}}
T_{\ell }\left( \varepsilon \right) \rho_d
\left( E^{* }-B_{i}-\varepsilon ,J_{d}\right),\label{Gamma}
\end{multline}%
where $J_{d}$ is the spin of the daughter nucleus, $S_{i}$, $J$, and $\ell $ are the spin, the total angular momentum, and the orbital angular momenta of the evaporated particle, $\varepsilon$ and $B_{i}$ are respectively its kinetic and separation energy, $T_{\ell }$ is its transmission coefficient or barrier penetration factor, and $\rho_d$ and $\rho _\mathrm{CN}$ are respectively the level density of the daughter and compound nucleus. The summations in Eq.~(\ref{Gamma}) include all angular momentum couplings between the initial and final states. In GEMINI, the Hauser-Feshbach formalism is implemented for the \textit{n}, \textit{p}, \textit{d}, \textit{t}, $^{3}$He, $\alpha $, $^{6}$He, $^{6-8}$Li, and $^{7-10}$Be channels.

The description of intermediate-mass fragment emission follows the Moretto formalism~\cite{Moretto:1975zz, Moretto:1988}, which has been further extended to the form of
\begin{eqnarray}
\Gamma _{\mathrm{Z}, \mathrm{A}}
&=&\frac{1}{2\pi \rho_\mathrm{CN}\left( E^{* },J_\mathrm{CN}\right) } \nonumber\\
&\times&  \int d\varepsilon \rho_\mathrm{sad}\left( E^* - B_{\mathrm{Z}, \mathrm{A}}(J_\mathrm{CN}) - \varepsilon ,J_\mathrm{CN}\right)
\end{eqnarray}
in GEMINI. In the above, $\rho_\mathrm{sad}$ is the level density at the saddle point, $\varepsilon$ is the kinetic energy in the fission degree of freedom at the saddle point, $B_{\mathrm{Z}, \mathrm{A}}(J_\mathrm{CN})$ is the conditional barrier depending on both the mass and charge asymmetries, and can be estimated as
\begin{eqnarray}
B_{\mathrm{Z}, \mathrm{A}}(J_\mathrm{CN})
 = B_{\mathrm{A}}^{Sierk}(J_\mathrm{CN}) + \Delta M + \Delta E_\mathrm{Coul}-\delta W - \delta P, \nonumber \\
\end{eqnarray}
where $\Delta M$ and $\Delta E_\mathrm{Coul}$ are the mass and Coulomb corrections that account for the different $Z$ and $A$ values of the two fragments, $\delta W$ and $\delta P$ are the ground-state shell and pairing corrections to the liquid drop barrier. Shell and pairing effects at the conditional saddle point are assumed to be small. The quantity $B_{\mathrm{A}}^{Sierk}$ is the interpolated Sierk barrier for the specified mass asymmetry. The total width requires summations over both the $Z$ and $A$ values of the lightest fragment.
For the symmetric divisions in heavy nuclei, the Bohr-Wheeler formalism~\cite{Bohr:1939ej} is used to predict the total symmetric fission yield
\begin{eqnarray}
\Gamma_\mathrm{BW} &=& \frac{1}{2 \pi \rho_\mathrm{CN}(E^*,J_\mathrm{CN})} \nonumber\\
        &\times& \int d\varepsilon\rho_\mathrm{sad}\left(E^*-B_f(J_\mathrm{CN})-\varepsilon,J_\mathrm{CN} \right),
\end{eqnarray}
where $B_f(J_\mathrm{CN})$ is the spin-dependent fission barrier, which is set to be the value from Sierk's finite-range model after the ground-state shell and pairing corrections, i.e.,
\begin{equation}
B_{f}(J_\mathrm{CN})= B_{f}^{Sierk}(J_\mathrm{CN})-\delta W - \delta P.
\end{equation}


\subsection{Calculation of the excitation energy}
\label{sec:testp}

The deexcitation of heavy clusters with $A \geq 4$ are handled by the GEMINI model~\cite{Charity:1988zz,Charity:2010wk}, which requires as inputs the angular momentum and the excitation energy of the cluster. The angular momentum of the cluster is calculated by summing those from all nucleons with respective to their C.M., while the energy of the cluster is calculated by averaging over $N_\mathrm{TP}$ parallel events for the same impact parameter and collision orientation based on a simplified SHF energy-density functional, i.e.,
\begin{eqnarray}\label{E}
E &=& \frac{1}{N_{TP}}\sum_{i} \left(\sqrt{m^2+p_i^2}-m\right) \notag\\
&+& \int d^3r \left[\frac{a}{2} \left(\frac{\rho}{\rho_0}\right)^2  +  \frac{b}{\sigma+1}\left(\frac{\rho}{\rho_0}\right)^{\sigma+1}\right] \notag\\
&+& \int d^3r E_{\mathrm{sym}}^{\mathrm{pot}} \left(\frac{\rho}{\rho_0}\right)^{\gamma} \frac{(\rho_n-\rho_p)^2}{\rho} \notag\\
&+& \int d^3r \left\{ \frac{G_S}{2} (\nabla \rho)^2 - \frac{G_V}{2} [\nabla (\rho_n-\rho_p)]^2\right\} \notag\\
&+& \frac{e^2}{2} \int d^3r d^3r' \frac{\rho_p(\vec{r})\rho_p(\vec{r}')}{|\vec{r}-\vec{r}'|} - \frac{3e^2}{4} \int d^3r \left( \frac{3\rho_p}{\pi}\right)^{4/3}. \notag\\
\end{eqnarray}
In the above, the first term represents the summation of the kinetic energy of the $i$th nucleon with $p_i$ and $m$ being its momentum and mass. The second term represents the isoscalar potential energy with $\rho(\vec{r})=\rho_n(\vec{r})+\rho_p(\vec{r})$ being the local nucleon number density at $\vec{r}$, $\rho_0=0.16$ fm$^{-3}$ being the saturation density, and parameters $a=-218$ MeV, $b=164$ MeV, and $\sigma=4/3$ chosen to reproduce empirical nuclear matter properties. The third term represents the potential part of the symmetry energy with $E_{\mathrm{sym}}^{\mathrm{pot}}=18$ MeV being its value at $\rho_0$, and $\gamma=0.3$ is used to describe the density dependence of the symmetry energy, which, however, has little effects on the final result. The fourth term represents the surface potential energy with empirical values of the isoscalar and isovector density gradient coefficients chosen to be $G_\mathrm{S}=132$ MeVfm$^5$ and $G_\mathrm{V}=5$ MeVfm$^5$~\cite{Chen:2010qx}. The fifth and sixth terms represent respectively the direct and exchange contribution of the Coulomb potential energy. The neutron and proton phase-space information are obtained from the test-particle method~\cite{Wong:1982zzb,Bertsch:1988ik}, i.e., calculated from averaging over parallel events for the same impact parameter and collision orientation, and this method is widely used in Boltzmann transport simulations of heavy-ion collisions~\cite{Xu:2019hqg}. Based on the same ground-state nucleon phase-space information, the binding energies per nucleon obtained from this method for most nuclei are found to agree with those from the SHFB calculation within $\pm1$ MeV, therefore justifying this approach. The excitation energy is then calculated by subtracting from the calculated cluster energy the ground-state energy of known nuclei taken from Ref.~\cite{Wang:2021xhn}. For clusters with compositions absent in Ref.~\cite{Wang:2021xhn}, an improved liquid-drop model~\cite{Wang:2014qqa} is employed to calculate their ground-state energies.

\subsection{Excitation energy due to electromagnetic field}
\label{sec:EBfield}

In relativistic heavy-ion collisions, the spectator matter is almost not affected by the participant matter, except that protons in the spectator matter can be affected by the retarded electromagnetic field generated by protons in another colliding nucleus. The increasing kinetic energy of protons is assumed to be dumped into total spectator matter, leading to the increase of the average excitation energy per nucleon there.

We use the Li\'enard-Wiechert formulaes to calculate the retarded electric and magnetic fields at the position $\vec{r}$ and time $t$, i.e.,
\begin{eqnarray}
\label{LWE}
e\vec{E}(t,\vec{r})&=&\frac{e^2}{4\pi \epsilon_0}\sum_i Z_i\frac{1-v_i^2}{(R_i-{\vec R}_i\cdot\vec{v}_i)^3}(\vec R_i-R_i\vec{v}_i),\nonumber\\
\\
\label{LWB}
e\vec{B}(t,\vec{r})&=&\frac{e^2}{4\pi \epsilon_0}\sum_i Z_i\frac{1-v_i^2}{(R_i-{\vec R}_i\cdot\vec{v}_i)^3}\vec{v}_i\times{\vec R}_i,
\end{eqnarray}
where $Z_i=1 (0)$ for protons (neutrons) is the charge number of the $i$th nucleon, ${\vec R}_i=\vec{r}-\vec{r}_i$ is the relative position of the field point $\vec{r}$ to the source point $\vec{r}_i$, and $\vec{r}_i$ is the location of the $i$th nucleon with velocity $\vec{v}_i$ at the retarded time $t_i=t-|\vec{r}-\vec{r}_i|/c$. The summations run over all charged particles in the system. Although there are singularities at $R_i=0$ in Eqs.~(\ref{LWE}) and (\ref{LWB}), such case does not exist in our calculation, since we only calculate the electromagnetic field for spectator protons generated by protons in another nucleus.

To calculate the electromagnetic field at moment $t$, we need to know the full phase-space information of all charged particles before $t$, and then calculate the retarded time $t_i$ for each source particle through the relation $t_i=t-|\vec{r}-\vec{r}_i|/c$.
We set the moment when the target and the projectile nuclei completely overlap as $t=0$, and assume that all nucleons are frozen in the rest frame of the nucleus, i.e., with their Fermi motion neglected. In the collision frame, the velocity and position of the $i$th nucleon can be expressed as
\begin{eqnarray}
\vec{v}_i&=&\left(0,0,\pm\sqrt{1-(2m_\mathrm{N}/\sqrtsnn)^2}\right), \\
\vec{r}_i &=& \vec{r}_i^0 + \vec{v}_it,
\end{eqnarray}
with $\vec{r}_i^0 = \vec{r}_i(t=0) = (x_i,y_i,2m_\mathrm{N}z_i/\sqrtsnn)$ being the position of the $i$th nucleon in the collision frame, given its position $(x_i,y_i,z_i)$ in the rest frame of the corresponding nucleus. We collect the phase-space information starting from $t=-5$ fm/c with the free propagating assumption of nucleons described above, in order to calculate the retarded electromagnetic field according to Eqs.~(\ref{LWE}) and (\ref{LWB}). The results are not much changed if we start at an even earlier time.

With the retarded electromagnetic field calculated according to Eqs.~(\ref{LWE}) and (\ref{LWB}) by the free propagating assumption of nucleons described above, we calculate the time evolution of the 4-momentum $p^\mu_i = (p^0_i, \vec{p}_i)$ of the $i$th nucleon in the collision frame under the electric and magnetic fields based on the equations of motion as follows
\begin{eqnarray}
\frac{\mathrm{d}p^0_i}{\mathrm{d}t}&=&eZ_i[\vec{E}(t,\vec{r}_i)+\vec{v}_i\times\vec{B}(t,\vec{r}_i)]\cdot \vec{v}_i,\\
\frac{\mathrm{d} \vec{p}_i}{\mathrm{d}t}&=&eZ_i[\vec{E}(t,\vec{r}_i)+\vec{v}_i\times\vec{B}(t,\vec{r}_i)].
\end{eqnarray}
The initial $(p^0_i, \vec{p}_i)$ in the collision frame is Lorentz boosted from that in the rest frame of the nucleus, while the final $(p^0_i, \vec{p}_i)$ in the collision frame is obtained by integrating from $t=-5$ fm/c to $t=5$ fm/c. We then Lorentz boost back $(p^0_i, \vec{p}_i)$ to the rest frame of the nucleus, and the increase of the kinetic energy from the difference in the initial and final $\vec{p}_i$ represents the excitation energy due to the electromagnetic field.

We note that the above approach is valid when the change of the momentum/energy of protons due to the electromagnetic field is very small and thus only a perturbation to the phase-space trajectory of source protons that generate the electromagnetic field. For a more self-consistent simulation of the particle motion and the electromagnetic field, see, e.g., Ref.~\cite{Ou:2011fm}.

\section{Results and discussions}
\label{sec:results}

With the theoretical framework described in detail above, we now discuss the relation between the free spectator nucleons and the neutron-skin thickness of colliding nuclei step by step, for the collision systems of \zrzr\ and \ruru\ at $\sqrtsnn=200$ GeV, \auau\ at $\sqrtsnn=130$ and 200 GeV, and \pbpb\ at $\sqrtsnn=5.02$ TeV. We first give the density distributions of relevant nuclei used in the present study. Then, we present the fitted $\nch$ distributions and the bulk properties of spectator matter using the Glauber model. Particularly, we study the production of free spectator nucleons and light clusters in different collision systems, and discuss possible probes of the neutron-skin thickness that can be measured experimentally. We also discuss the effect of the electromagnetic field on the excitation energy of heavy fragments and the possible impact on the relevant results.

\subsection{Density and $\nch$ distributions}

\begin{figure}[!h]
\includegraphics[width=1\linewidth]{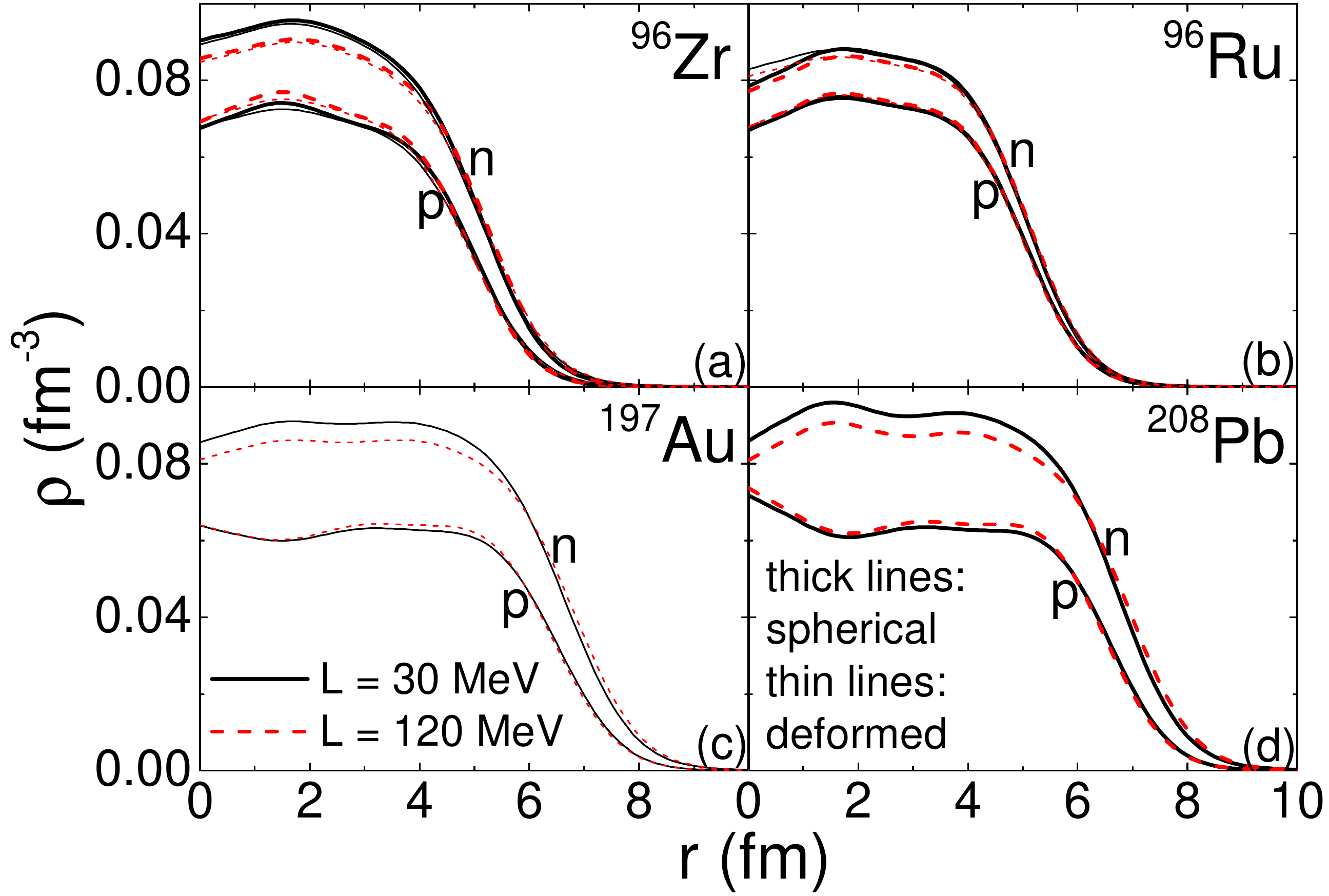}
\vspace*{-.3cm}
\caption{\label{fig:density} Density profiles for neutrons (n) and protons (p) in $^{96}$Zr, $^{96}$Ru, $^{197}$Au, and $^{208}$Pb from SHFB calculations using different slope parameters $L$ of the symmetry energy. Thick lines are from calculations based the assumption of spherical nuclei, and thin lines are averaged profiles over orientations from calculations with $\beta_{2}=0.06$ and $\beta_3=0.2$ for $^{96}$Zr, $\beta_2=0.16$ for $^{96}$Ru, and $\beta_2=-0.15$ for $^{197}$Au, respectively.}
\end{figure}

\begin{table}[h!]
\centering
\caption{Neutron-skin thicknesses $\Delta r_\mathrm{np}$ and deformation parameters $\beta_{2}$ and $\beta_{3}$ for different nuclei using different slope parameters $L$ of the symmetry energy from SHFB calculations.}
\label{tab:SHF}
\renewcommand\arraystretch{1.5}
\setlength{\tabcolsep}{2mm}
\begin{tabular}{|c|c|c|c|}
\hline
\multirow{2}{*}{Nucleus}    & \multirow{2}{*}{$\beta_2,~\beta_3$}&   \multicolumn{2}{c|}{$\Delta r_\mathrm{np}$(fm)}  \\
\cline{3-4}
                            &                           &   {$L=30$ MeV}  & {$L=120$ MeV } \\
\hline
\multirow{2}{*}{$^{96}$Zr}  & 0, 0                         & 0.147 & 0.231 \\
                            & 0.06, 0.2~\cite{Zhang:2021kxj} & 0.145 & 0.227 \\
\hline
\multirow{2}{*}{$^{96}$Ru}  & 0, 0                         & 0.028 &  0.061\\
                            & 0.16, 0~\cite{Zhang:2021kxj} & 0.026 & 0.058 \\
\hline
$^{197}$Au                  & -0.15, 0~\cite{Giacalone:2021udy,Moller:2015fba}  & 0.127 & 0.243\\
\hline
$^{208}$Pb                  & 0, 0                         &  0.149 & 0.281 \\
\hline
\end{tabular}
\end{table}

\begin{figure}[!h]
\includegraphics[width=1\linewidth]{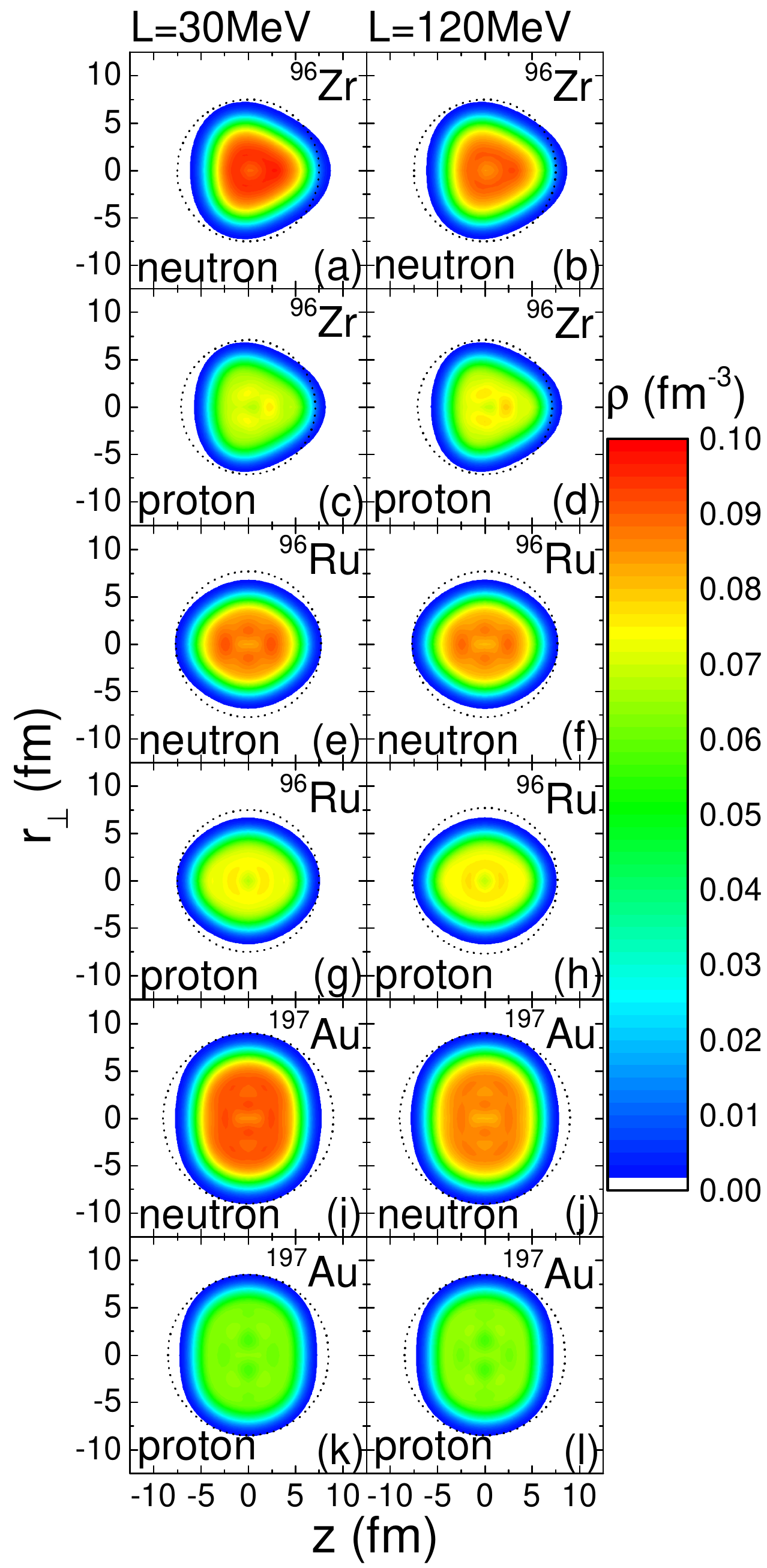}
\vspace*{-.3cm}
\caption{\label{fig:density_con} Density contours of neutrons and protons in the $r_\perp-z$ plane for $^{96}$Zr with $\beta_2=0.06$ and $\beta_3=0.2$, $^{96}$Ru with $\beta_2=0.16$, and $^{197}$Au with $\beta_2=-0.15$, respectively, from deformed SHFB calculations using $L=30$ (left) and 120 (right) MeV. Dotted lines are spherical curves for the comparison with deformed shapes.}
\end{figure}

The advantage of obtaining the nucleon density distribution from SHFB calculations is that the resulting neutron-skin thickness can be consistently calculated from a well-defined energy-density functional with a given slope parameter $L$ of the symmetry energy, different from a parameterized density distribution, e.g., a spherical or deformed Woods-Saxon form. Figure~\ref{fig:density} shows the density profiles for neutrons and protons in $^{96}$Zr, $^{96}$Ru, $^{197}$Au, and $^{208}$Pb from SHFB calculations with different $L$, where results for deformed nuclei are averaged over orientations. One sees that the density distribution of neutrons is more diffusive with a larger $L$, but this is opposite to that of protons, resulting in a larger neutron-skin thickness $\Delta r_\mathrm{np}$ with a larger $L$, especially for a more neutron-rich nucleus. The averaged density profiles over orientations for deformed $^{96}$Zr and $^{96}$Ru are similar to those of a spherical shape. The $\Delta r_{\mathrm{np}}$ in different nuclei with different deformation parameters and $L$ are listed in Table~\ref{tab:SHF}. We can see that $\Delta r_{\mathrm{np}}$ can be varied by only less than $8\%$ for the same $L$ with different deformation parameters. For completeness, we also display the deformed density contours for neutrons and protons in the $r_\perp-z$ plane of $^{96}$Zr with $\beta_{2}=0.06$ and $\beta_{3}=0.2$, $^{96}$Ru with $\beta_2=0.16$, and $^{197}$Au with $\beta_2=-0.15$ from SHFB calculations using different $L$ in Fig.~\ref{fig:density_con}, where the $z$ axis is the orientation of the symmetric axis and $r_\perp$ is perpendicular to $z$. The 2-dimensional density distributions are of course consistent with the density profiles shown in Fig.~\ref{fig:density} but including more information, such as the density oscillations inside the nucleus and the slightly different deformation for neutrons and protons. We find that the deformation parameters $\beta_2$ and $\beta_3$ for the distribution of neutrons can be different from that of protons within about $10\%$.

\begin{figure}[!h]
\includegraphics[width=0.8\linewidth]{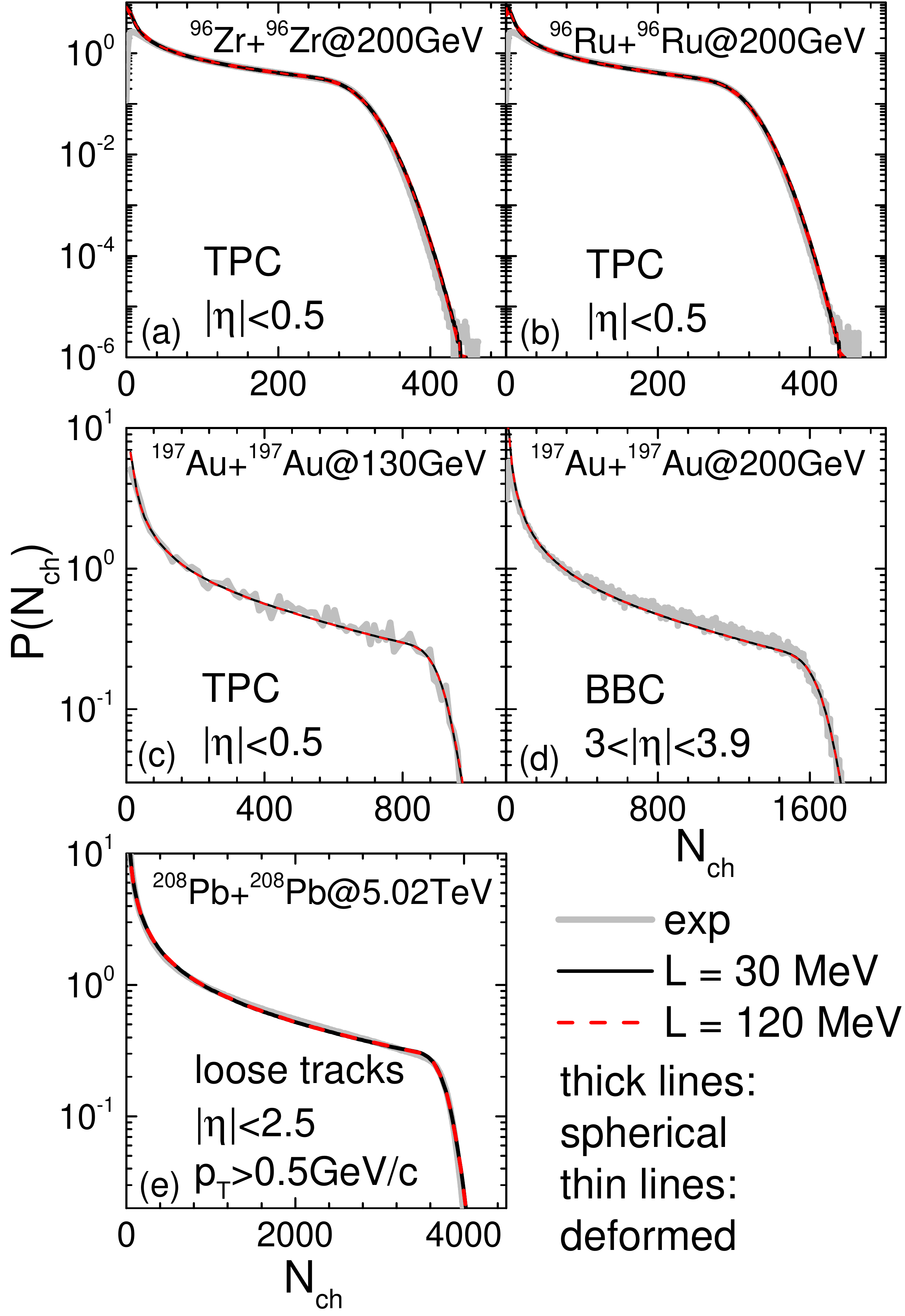}
\vspace*{-.3cm}
\caption{\label{fig:nch} Rescaled distributions of charged-particle multiplicity $\nch$ in different collision systems, where results from the two-component Glauber model with the fitted parameters listed in Table~\ref{tab:glauberpara} are compared with experimental data~\cite{STAR:2021mii,STAR:2000ekf,PHENIXBBC200,ATLAS:2022dov}. The fitted results by the two-component Glauber model are based on density distributions for $^{96}$Zr with $\beta_{2(3)}=0(0)$ and 0.06(0.2), $^{96}$Ru with $\beta_2=0$ and 0.16, $^{197}$Au with $\beta_2=-0.15$, and $^{208}$Pb with $\beta_2=0$, obtained from SHFB calculations.}
\end{figure}

\begin{table}[h!]
\centering
\caption{Values of parameters in the two-component Glauber model for fitting the charged-particle multiplicity distribution in different collision systems.}
\label{tab:glauberpara}
\renewcommand\arraystretch{1.5}
\setlength{\tabcolsep}{3.5mm}
\begin{tabular}{ccccc}
\hline\hline
       & $\sqrt{s_\mathrm{NN}}$      & $x$ & $\bar{n}$ & $m$  \\
\hline
 \zrzr\ & 200 GeV& 0.12  & 2.3  & 2.0  \\
\hline
 \ruru\ & 200 GeV& 0.12  & 2.3  & 2.2  \\
\hline
 \auau\ & 130 GeV&  0 &  4.8 & 4.6  \\
\hline
 \auau\ & 200 GeV&  0.10 &  5.8 & 2.3  \\
\hline
 \pbpb\ & 5.02 TeV& 0.09  & 10.3 & 3.2  \\
\hline\hline
\end{tabular}
\end{table}

With the density distributions presented above, we fit the distributions of charged-particle multiplicity $\nch$ in collision systems of \zrzr\ and \ruru\ at $\sqrtsnn=200$ GeV, \auau\ at $\sqrtsnn=130$ and 200 GeV, and \pbpb\ at $\sqrtsnn=5.02$ TeV with the two-component Glauber model, and the results are shown in Fig.~\ref{fig:nch}, where the corresponding experimental data are also plotted for comparison. $\nch$ in Fig.~\ref{fig:nch} (c) near the mid-pseudorapidity region measured by the STAR Collaboration~\cite{STAR:2000ekf} is rescaled with a different $\nch^\mathrm{max}$ so that comparable to the $\nch$ range at large pseudorapidities measured by the PHENIX Collaboration~\cite{PHENIX}. The parameters of the two-component Glauber model in Eqs.~(\ref{N_s}) and (\ref{pnbd}) for fitting the charged-particle distribution in different collision systems are listed in Table~\ref{tab:glauberpara}, and the parameter values are almost the same for nucleon distributions obtained with different $L$ or deformation parameters from SHFB calculations.

\subsection{Isospin properties of spectator matter}

\begin{figure}[!h]
\includegraphics[width=1\linewidth]{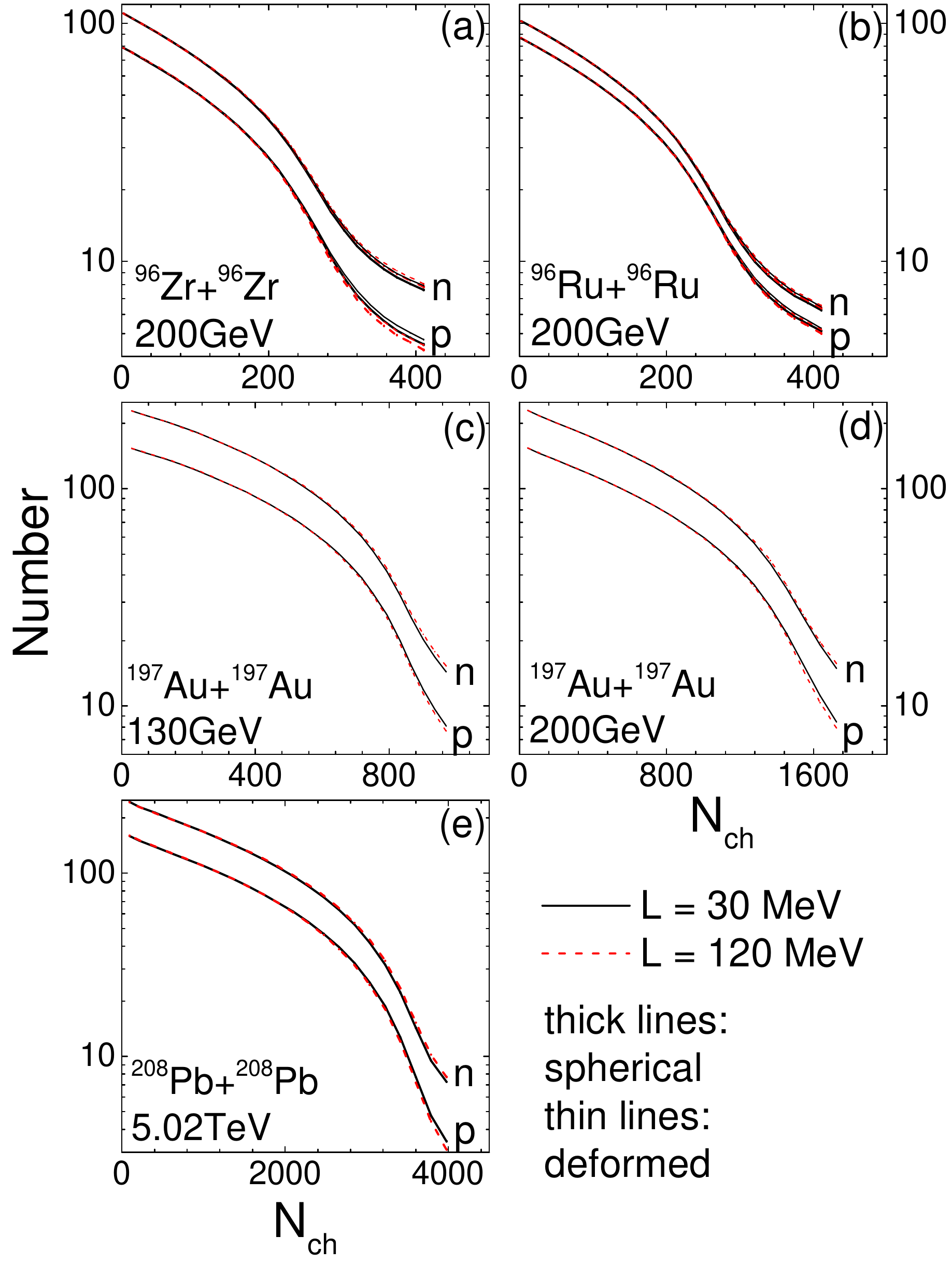}
\vspace*{-.3cm}
\caption{\label{fig:total_spectator_nch} Numbers of total spectator neutrons (n) and protons (p) as a function of $\nch$ in different collision systems from the Glauber model with density distributions from SHFB calculations using a smaller or larger $L$ and with $\beta_{2(3)}=0(0)$ and 0.06(0.2) for $^{96}$Zr, $\beta_2=0$ and 0.16 $^{96}$Ru, $\beta_2=-0.15$ for $^{197}$Au, and $\beta_2=0$ for $^{208}$Pb.}
\end{figure}

\begin{figure}[ht]
\includegraphics[width=1\linewidth]{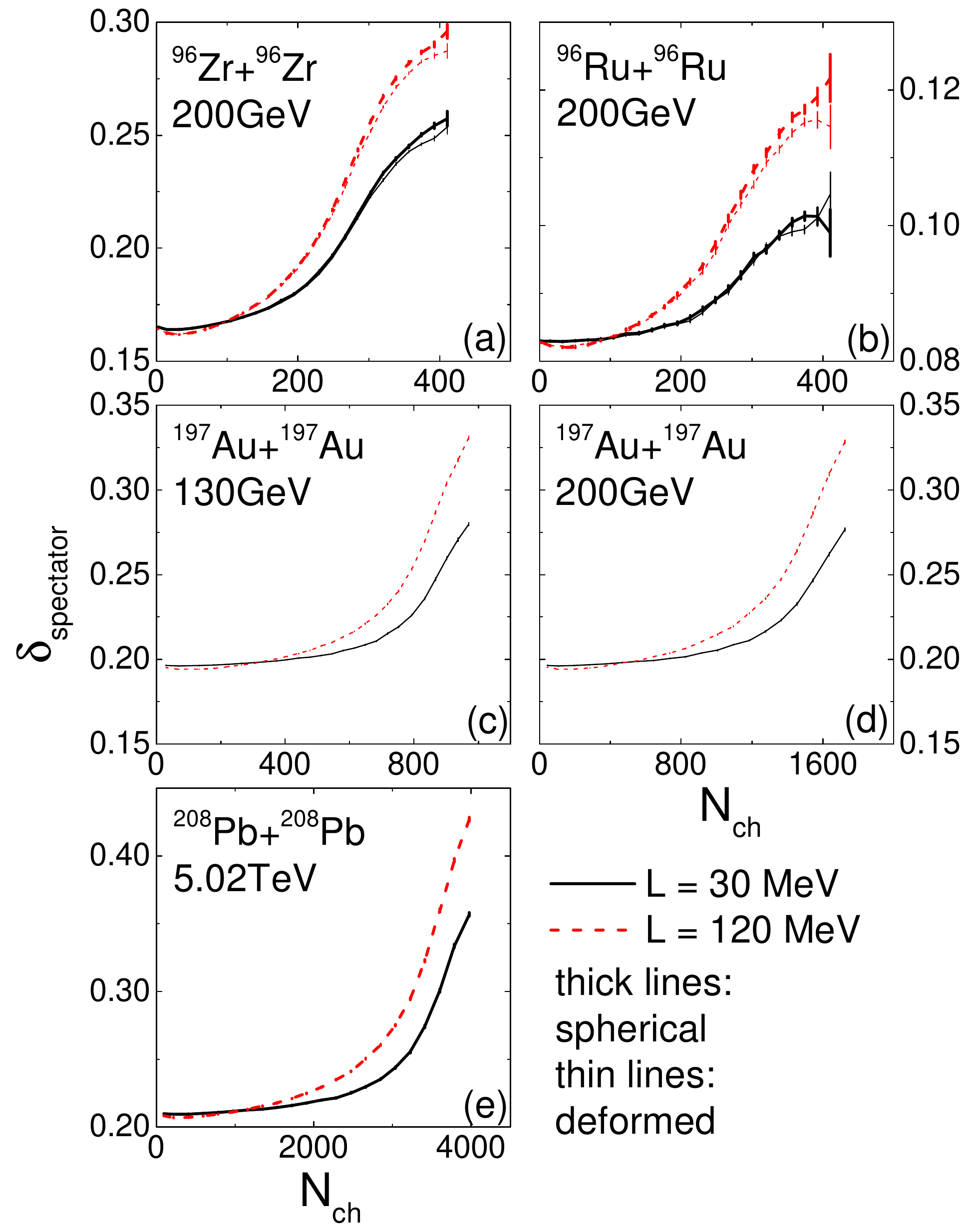}
\caption{\label{fig:isospin_asy_nch} Isospin asymmetries of spectator matter as a function of $\nch$ in different collision systems from the Glauber model with density distributions from SHFB calculations using a smaller or larger $L$ and with $\beta_{2(3)}=0(0)$ and 0.06(0.2) for $^{96}$Zr, $\beta_2=0$ and 0.16 $^{96}$Ru, $\beta_2=-0.15$ for $^{197}$Au, and $\beta_2=0$ for $^{208}$Pb.}
\end{figure}

In the present study, the dynamics of the participant matter is neglected, and we only consider the spectator matter from the Glauber model at different impact parameters, or for different $\nch$ as in the experimental analysis. The spectator matter is expected to be more neutron-rich for colliding nuclei with a larger overall isospin asymmetry or a larger $\Delta r_\mathrm{np}$, especially at large $\nch$ or $\npart$. The numbers of spectator neutrons and protons as a function of $\nch$ in the collision systems of \zrzr\ and \ruru\ at $\sqrtsnn=200$ GeV, \auau\ at $\sqrtsnn=130$ and 200 GeV, and \pbpb\ at $\sqrtsnn=5.02$ TeV are shown in Fig.~\ref{fig:total_spectator_nch}. Both numbers decrease with increasing $\nch$, corresponding to less spectator nucleons at smaller centralities. The difference between results from different deformation parameters are not visible, while the symmetry energy effect is only visible in the UCC region with a large $\nch$, where the number of neutrons (protons) is larger with a larger (smaller) value of $L$.

To further demonstrate the isospin properties of the spectator matter with different symmetry energies, we display in Fig.~\ref{fig:isospin_asy_nch} the overall isospin asymmetry $\delta_{\rm spectator}=(N-Z)/(N+Z)$ of the spectator matter as a function of $\nch$, where $N$ and $Z$ are the total neutron and proton number in the spectator matter, respectively. Due to the presence of the neutron skin, the spectator matter becomes more neutron-rich in more central collisions, so that the results in the UCC region are generally more sensitive to the neutron skin, thus serving as a good probe of the $\Delta r_{\mathrm{np}}$ and $L$. Inclusion of the nuclear deformation reduces slightly the $\delta_{\mathrm{spectator}}$, since the nuclear deformation, combined with random collision orientation, smears the radial distribution of the spectator protons and neutrons and generates slightly more spectator nucleons than that in the spherical case. But such effect is small compared to the influence of $L$. We note that the results at large $\nch$ is already smeared, since there is a distribution of $\nch$ for a given $\npart$ or impact parameter. If we plot the isospin asymmetry of spectator matter as a function of $\npart$ or impact parameter, $\delta_{\rm spectator}$ can be even larger in the UCC region.


\subsection{Spectator particle yield}

\begin{figure*}[h]
\includegraphics[width=0.45\linewidth]{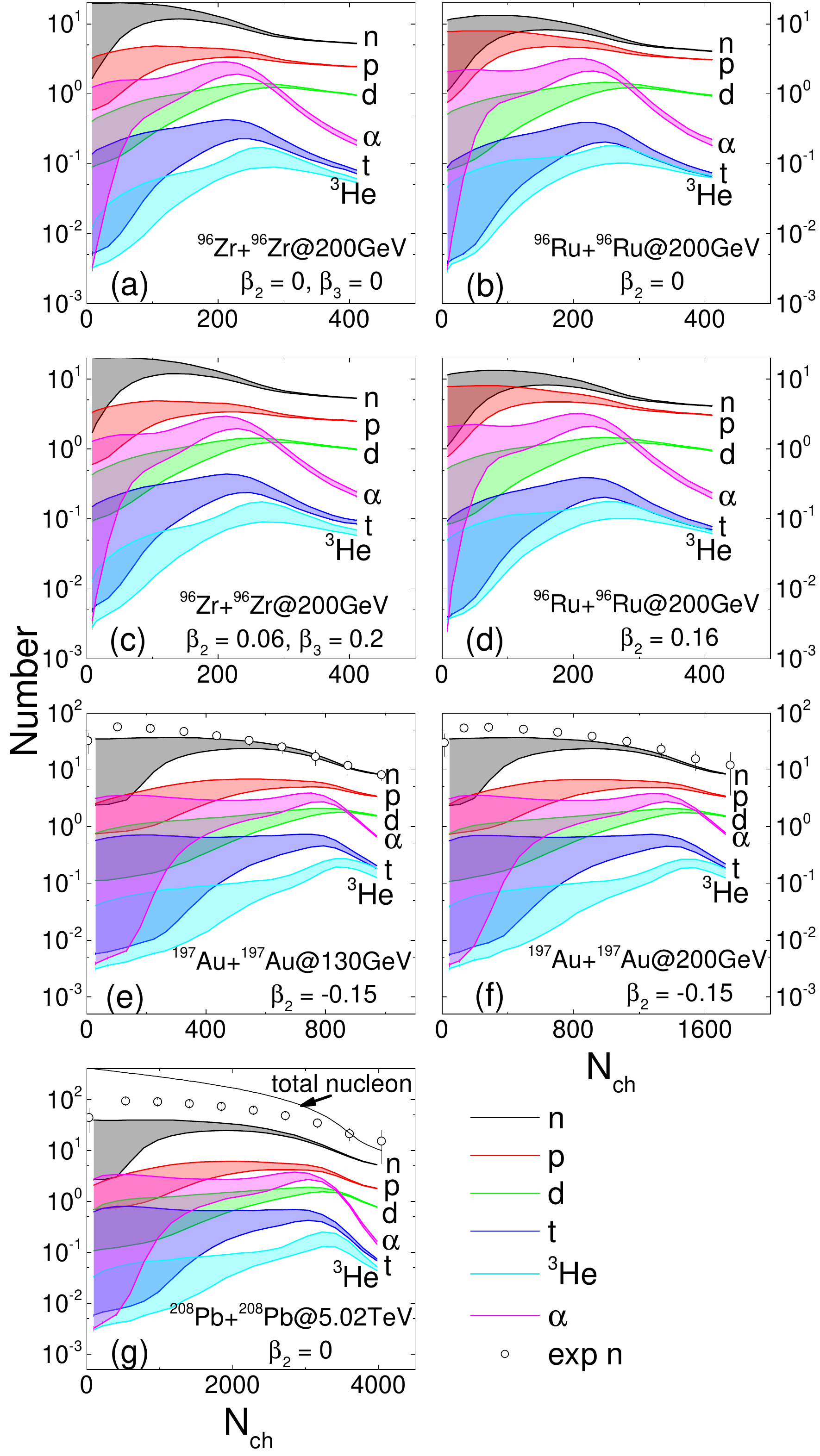}
\includegraphics[width=0.45\linewidth]{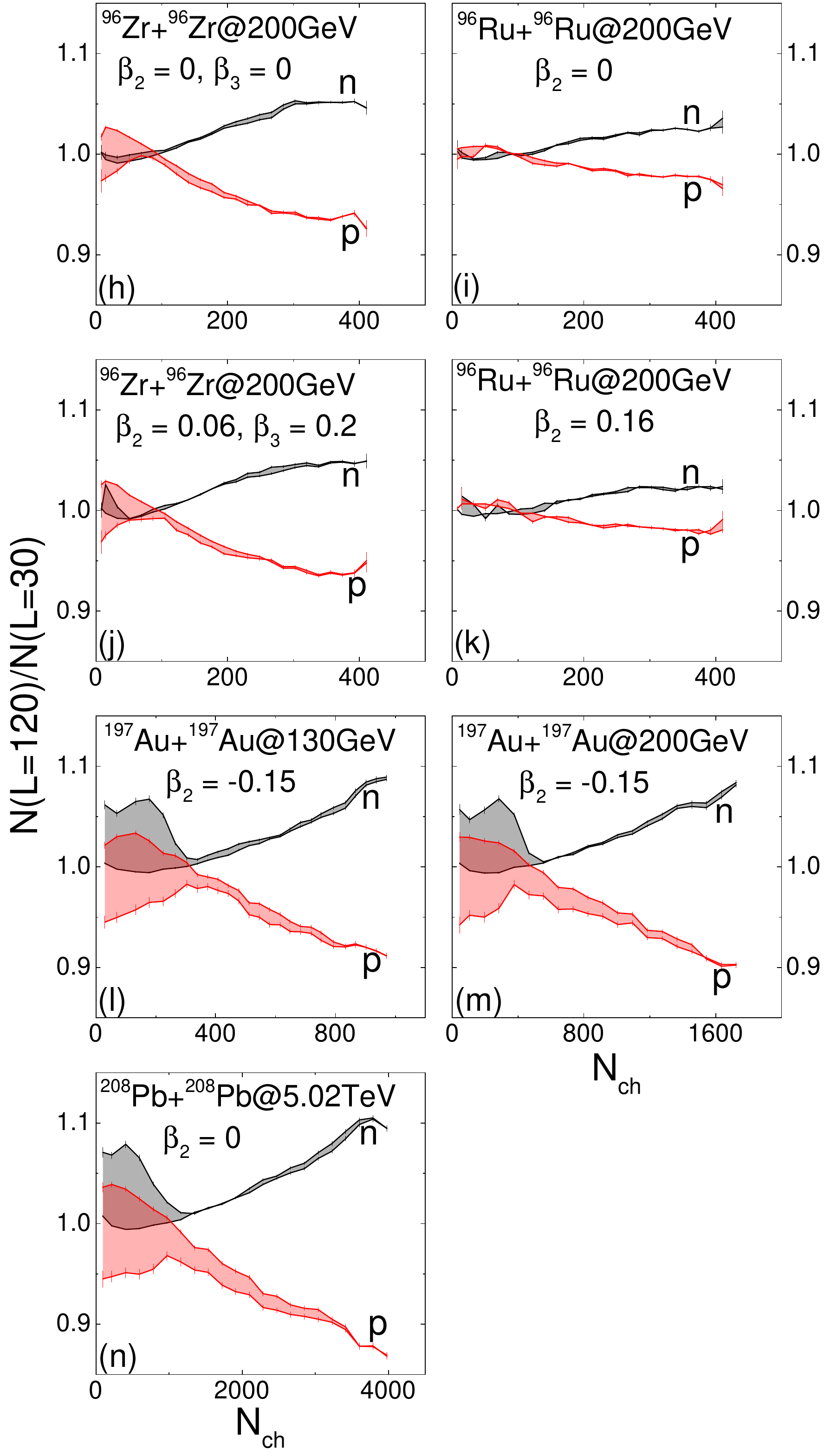}
\vspace*{-.3cm}
\caption{\label{fig:yield} Left: Numbers of free spectator nucleons and light clusters from \zrzr\ at $\sqrtsnn=200$ GeV, \ruru\ at $\sqrtsnn=200$ GeV, \auau\ at $\sqrtsnn=130$ GeV, \auau\ at $\sqrtsnn=200$ GeV, and \pbpb\ at $\sqrtsnn=5.02$ TeV collision systems for $L=30$ MeV and with different deformation parameters. Estimated spectator neutrons from experimental data measured by ZDC in the collisions of \auau\ at $\sqrtsnn=130$ GeV~\cite{PHENIX}, \auau\ at $\sqrtsnn=200$ GeV~\cite{PHENIX:2003iij}, and \pbpb\ at $\sqrtsnn=5.02$ TeV~\cite{ATLAS:2781150,ALICE:2013hur} are also plotted for comparison. Right: Yield ratios of free spectator neutrons and protons for $L=120$ MeV to those for $L=30$ MeV in the corresponding systems as in the left panels. Bands represent the uncertainty of $\pm 1$ MeV per nucleon in calculating the excitation energy for the deexcitation of heavy fragments by GEMINI.}
\end{figure*}

The isospin-dependent property of spectator matter may manifest itself in the particle yield from the direct production and fragmentation of the spectator matter. The left panels of Fig.~\ref{fig:yield} display the numbers of final free particles of different species, including neutrons, protons, deuterons, tritons, $^3$He, and $\alpha$ particles, in different collision systems and based on initial density distributions obtained by SHFB calculations using $L=30$ MeV. For free spectator neutrons and protons, they are composed of the residue ones from direct production that have not coalesced into light clusters and those from the deexcitation of heavy clusters by GEMINI. For spectator deutrons, tritons, and $^3$He, they are formed from the coalescence of direct produced neutrons and protons as well as from the deexcitation of heavy clusters by GEMINI. For $\alpha$ particles or even heavier clusters, they are mostly produced from the deexcitation of heavy clusters by GEMINI as well as the stable ones from the clusterization algorithm. The bands in Fig.~\ref{fig:yield} represent the uncertainty of $\pm 1$ MeV per nucleon in calculating the excitation energy for the deexcitation of heavy fragments by GEMINI, estimated from the same degree of accuracy for calculating the ground-state energy of the relevant nucleus from the test-particle method compared to the SHFB calculation. The uncertainties from the deexcitation process are seen to be largely reduced, especially for neutrons and protons with the largest multiplicities, in the UCC region where there are very few heavy clusters. One observes a larger multiplicity for free spectator neutrons than protons, particularly in a more neutron-rich collision system (e.g., \zrzr) than in a less neutron-rich  collision system (e.g., \ruru).

It is interesting to see that the multiplicities of most particle species displayed here first increase and then decrease with increasing $\nch$, different from the monotonic decreasing trend of the total nucleon numbers in spectator matter observed in Fig.~\ref{fig:total_spectator_nch}. This is because in the ultraperipheral collision region, more nucleons are bounded in stable heavy clusters rather than existing as free nucleons or light clusters shown here. One sees that the multiplicities for tritons, $^3$He, and $\alpha$ particles are larger in \auau\ collisions compared to that in \zrzr\ and \ruru\ collisions, likely due to more nucleons in the heavier \auau\ collision system. On the other hand, although the multiplicities of different particle species are almost similar in \auau\ collisions at $\sqrtsnn=130$ and 200 GeV, there are fewer spectator tritons, $^3$He, and $\alpha$ particles at LHC energies compared to those at RHIC energies. This is likely due to the similar NN cross section at $\sqrtsnn=130$ and 200 GeV, but a quite large NN cross section at $\sqrtsnn=5.02$ TeV, as shown in Table.~\ref{tab:crosssection}.

The comparisons with the ZDC data in \auau\ and \pbpb\ collisions need some further discussions. Using the clusterization parameter $\Delta r_{\rm max}=3$ fm, we reproduce the ZDC data in \auau\ collisions at $\sqrtsnn=130$ GeV for $\nch >700$, while we underestimate the ZDC data at smaller $\nch$ in the same collision system, likely due to other mechanisms at smaller centralities. Although the NN cross section at $\sqrtsnn=200$ GeV is only 2 mb larger than that at $\sqrtsnn=130$ GeV as shown in Table.~\ref{tab:crosssection}, which leads to almost the same total spectator nucleon number in Au+Au collisions at $\sqrtsnn=130$ and 200 GeV as seen from Fig.~\ref{fig:total_spectator_nch} (c) and (d), we underestimate the ZDC data in ultracentral \auau\ collisions at $\sqrtsnn=200$ GeV based on the same approach. In addition, as shown in Fig.~\ref{fig:yield} (g), the total spectator nucleon number obtained based on the Glauber model is seen to be smaller than the ZDC data in ultracentral \pbpb\ collisions at $\sqrtsnn=5.02$ TeV. The apparent differences between the experimental data and our predictions in the UCC region at higher collision energies could point to possible background particles from the interaction region that reach ZDC acceptance. A recent calculation by the Abrasion-Ablation Monte Carlo model for colliders~\cite{Nepeivoda:2022dvc} also significantly underpredicts the ZDC data for neutrons from the ALICE Collaboration~\cite{ALICE}, suggesting background contributions in the UCC region. We estimate the possible background contribution using the widely used AMPT model~\cite{Lin:2004en}. Within the pseudorapidity acceptance of ZDC, $|\eta|>5.9$ for \auau\ collisions at $\sqrtsnn=130$ and 200 GeV in PHENIX and $|\eta|>8.3$ for \pbpb\ collisions at $\sqrtsnn=5.02$ TeV in ATLAS, we observe considerable amount of charge-neutral and long-lived particles, such as $\pi^0$, antineutrons, $K_0^{\rm long}$, photons, and also neutrons from the participant region. Even if the energies of these particles are transformed to equivalent neutron numbers, which is closer to how ZDC works, the background contribution is still not small. The impact of these particles should be properly estimated and subtracted, before the ZDC data can be directly compared to the theoretical results in the present study.

We have also displayed the yield ratios of free spectator neutrons and protons for $L=120$ MeV to those for $L=30$ MeV in the right panels of Fig.~\ref{fig:yield}. The uncertainties of the deexcitation process, represented by the bands as in the left panels, are more obviously seen in collision systems with heavier nuclei, as a result of more heavy fragments. On the other hand, the increasing (decreasing) trend of the yield ratio for neutrons (protons) with increasing $\nch$ is seen in all collision systems, especially in more neutron-rich collision systems, consistent with the larger $\delta_{\mathrm{spectator}}$ for larger $L$ and in more neutron-rich systems shown in Fig.~\ref{fig:isospin_asy_nch} in the UCC region. For the yield ratios of other particles shown in the left panels of Fig.~\ref{fig:yield} for $L=120$ MeV to those for $L=30$ MeV, we found that they are closer to 1 and less sensitive to $\nch$, compared to those of neutrons and protons.

\subsection{Probing neutron-skin thickness}

\begin{figure}[!h]
\includegraphics[width=1\linewidth]{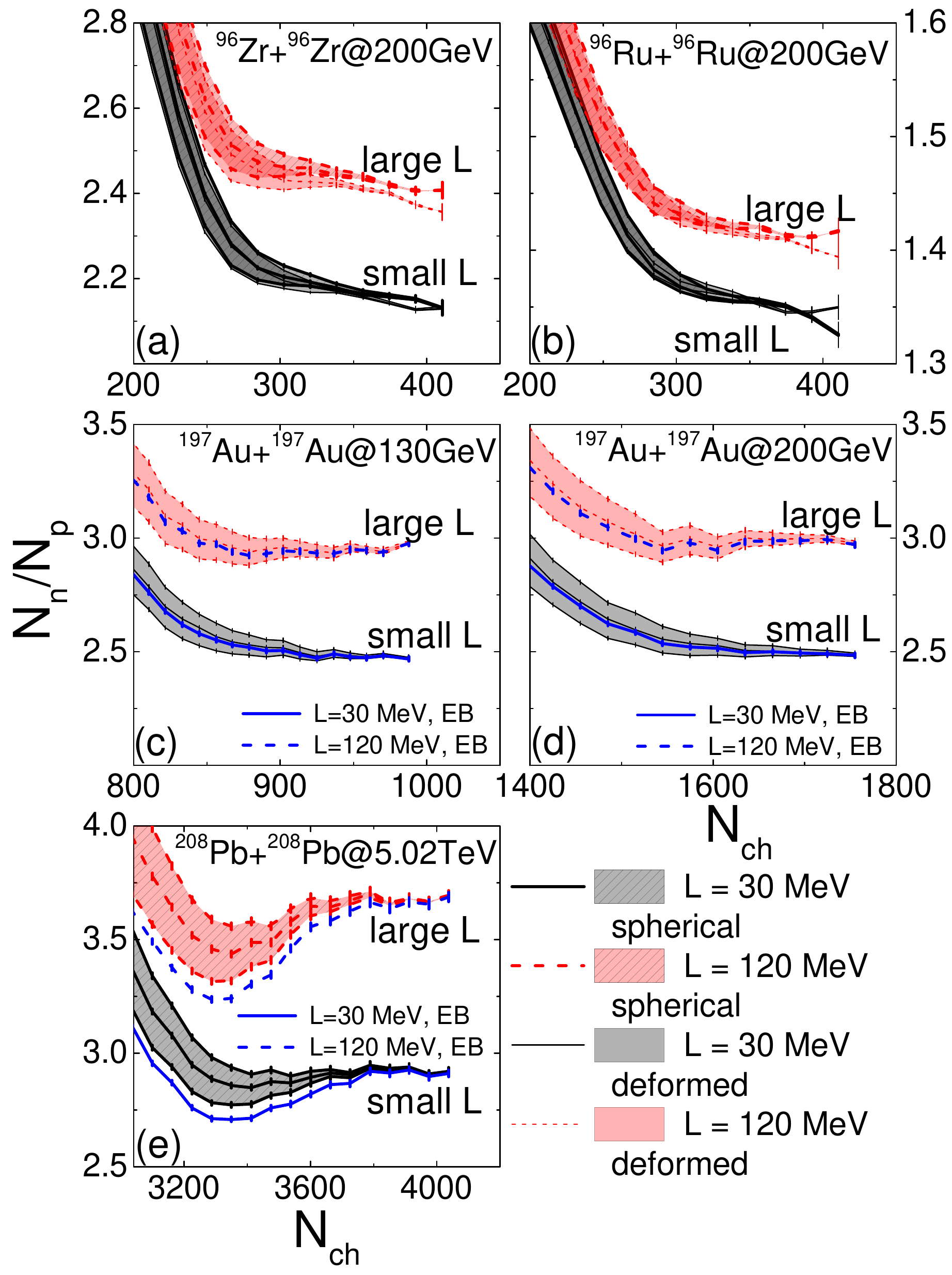}
\vspace*{-.3cm}
\caption{\label{fig:ratio} Yield ratio $N_n/N_p$ of free spectator neutrons to protons for different scenarios and in different collision systems. Bands represent the uncertainty of $\pm 1$ MeV per nucleon in calculating the excitation energy for the deexcitation of heavy fragments by GEMINI. For \auau\ collisions at $\sqrtsnn=130$ and 200 GeV and \pbpb\ collisions at $\sqrtsnn=5.02$ TeV, results after considering the excitation by the electromagnetic field (EB) are also plotted for comparison.}
\end{figure}

\begin{table}[h!]
\centering
\caption{Yield ratio $N_n/N_p$ of free spectator neutrons to protons in the UCC region of different collision systems and for different slope parameters $L$ (in MeV) of the symmetry energy.}
\label{tab:n/p}
\renewcommand\arraystretch{1.5}
\setlength{\tabcolsep}{1.2mm}
\begin{tabular}{|c|c|c|c|c|}
\hline
\multirow{2}{*}{} & \multirow{2}{*}{\sqrtsnn} & \multirow{2}{*}{$\beta_2,~\beta_3$}&   \multicolumn{2}{c|}{$N_n/N_p$}  \\
\cline{4-5}
                                    &           &                &   {$L=30$}  & {$L=120$} \\
\hline
\multirow{2}{*}{\zrzr} & \multirow{2}{*}{200 GeV}   & 0, 0                         & 2.15 & 2.40 \\
                            &        & 0.06, 0.2                      & 2.15 & 2.35 \\
\hline
\multirow{2}{*}{\ruru} & \multirow{2}{*}{200 GeV}   & 0, 0                         & 1.35 & 1.40 \\
                            &        & 0.16, 0                      & 1.35 & 1.40 \\
\hline
\auau & 130 GeV                    & -0.15, 0                     &  2.5 & 3.0 \\
\hline
\auau & 200 GeV                    & -0.15, 0                     &  2.5 & 3.0 \\
\hline
\pbpb & 5.02 TeV                   & 0, 0                         &  2.9 & 3.7 \\
\hline
\end{tabular}
\end{table}

\begin{figure}[!h]
\includegraphics[width=0.8\linewidth]{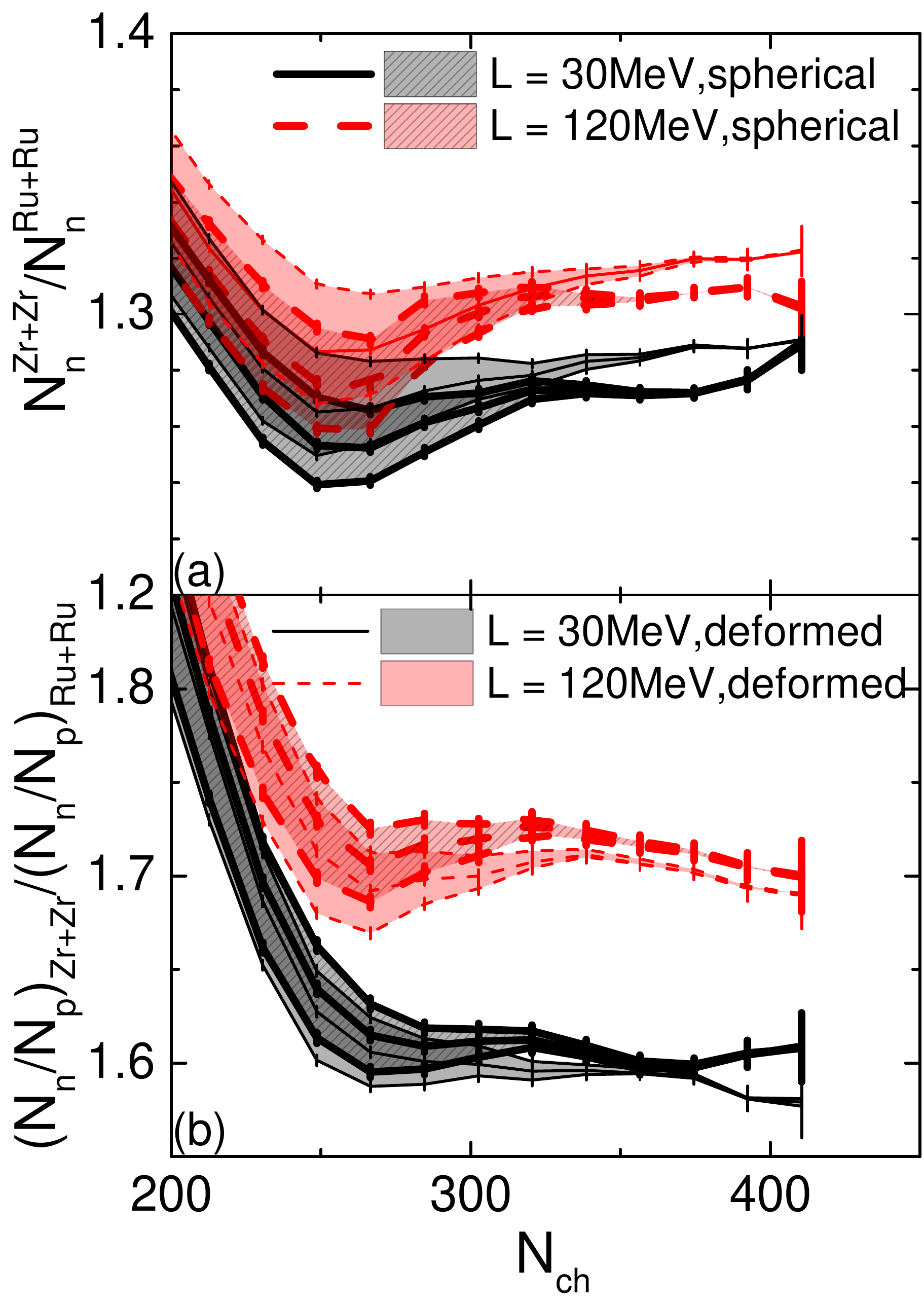}
\vspace*{-.3cm}
\caption{\label{fig:ratio2} The ratio of $N_n$ (upper) and $N_n/N_p$ (lower) in \zrzr\ to \ruru\ collisions at $\sqrtsnn=200$ GeV based on density distributions of $^{96}$Zr and $^{96}$Ru using different $L$ and deformation parameters from SHFB calculations. Bands represent the uncertainty of $\pm 1$ MeV per nucleon in calculating the excitation energy for the deexcitation of heavy fragments by GEMINI.}
\end{figure}

The number of free spectator particles presented above suffer from the theoretical uncertainties, such as the clusterization algorithm and the deexcitation process, etc., while taking the ratio of particle yields generally reduces these uncertainties, since the yields of different particle species are generally underestimated or overestimated simultaneously due to theoretical or experimental uncertainties. The yield ratio of neutrons to protons is one of the most sensitive probes of the symmetry energy in low-energy heavy-ion collisions (see, e.g., Ref.~\cite{Li:1997rc}), and we propose here that the yield ratio of free spectator neutrons to protons $N_n/N_p$ as a sensitive probe of the neutron-skin thickness of the colliding nuclei. As shown in the right panels of Fig.~\ref{fig:yield}, the symmetry energy or the neutron-skin thickness has opposite effect on $N_n$ and $N_p$, so taking the yield ratio enhances the effect compared to the yield of a single particle species. Figures~\ref{fig:ratio} displays the $N_n/N_p$ at small centralities in different collision systems and in different scenarios. One sees that the overall ratio is larger in more neutron-rich collision systems, consistent with the total isospin asymmetry $\delta_{\rm spectator}$ of spectator matter shown in Fig.~\ref{fig:isospin_asy_nch}. The bands represent the uncertainty of $\pm 1$ MeV per nucleon in calculating the excitation energy for the deexcitation of heavy fragments by GEMINI as in Fig.~\ref{fig:yield}. A higher excitation energy generally leads to more free nucleons and a smaller $N_n/N_p$ ratio, and the ratio at extremely large $\nch$ is shown to be independent of such uncertainties since the deexcitation of heavy clusters is unimportant for the production of free nucleons there. The difference of the $N_n/N_p$ ratio at extremely large $\nch$ is larger for a larger $\Delta r_\mathrm{np}$ or $L$, and thus a robust probe for them. Considering the difference between $L=30$ and 120 MeV, the $N_n/N_p$ ratio changes from about 2.9 to 3.7 in ultracentral \pbpb\ collisions at $\sqrtsnn=5.02$ TeV, from about 2.5 to 3 in ultracentral \auau\ collisions at both $\sqrtsnn=200$ and 130 GeV. Assuming both $^{96}$Zr and $^{96}$Ru are spherical, the $N_n/N_p$ ratio changes from about 2.15 to 2.40 in ultracentral \zrzr\ collisions, and from about 1.35 to 1.40 in ultracentral \ruru\ collisions, for $L=30$ and 120 MeV, respectively. Considering the deformation of $^{96}$Zr and $^{96}$Ru, the $N_n/N_p$ ratios are reduced by less than 0.05 in ultracentral \zrzr\ and \ruru\ collisions. The above ratios are listed in Table~\ref{tab:n/p}. It is seen that the effect of $L$ is about $28\%$ in \pbpb\ collisions, $20\%$ in \auau\ collisions, $12\%$ in \zrzr\ collisions, and even smaller in \ruru\ collisions. It is thus recommended to probe $\Delta r_\mathrm{np}$ or $L$ from the $N_n/N_p$ ratio in a heavier and more neutron-rich system, if both $N_n$ and $N_p$ can be measured accurately in a single collision system. The smaller $N_n/N_p$ ratios for the deformed case are consistent with the smaller $\delta_{\mathrm{spectator}}$ shown in Fig.~\ref{fig:isospin_asy_nch} and a little smaller $\Delta r_\mathrm{np}$ shown in Table.~\ref{tab:SHF}, while the deformation is just a secondary effect compared to that from $\Delta r_\mathrm{np}$ or $L$ and controllable once the deformation is known.

Taking the ratio of observables in isobaric collision systems is also an effective way of reducing uncertainties from both theoretical and experimental sides. Motivated by a significant enhancement of $N_n$ in \zrzr\ collisions relative to \ruru\ collisions measured by the ZDC shown in the recent STAR paper~\cite{STAR:2021mii}, we propose that the ratio of $N_n$ in the isobaric collision systems is a useful probe of the $\Delta r_\mathrm{np}$ of colliding nuclei and the slope parameter $L$ of the symmetry energy~\cite{Liu:2022kvz}. As shown in the upper panel of Fig.~\ref{fig:ratio2}, the ratio of $N_n$ in ultracentral collisions is free from the uncertainties of the deexcitation process, with the latter indicated again by the bands as in Figs.~\ref{fig:yield} and \ref{fig:ratio}. Assuming that both $^{96}$Zr and $^{96}$Ru are spherical, the $N_n$ ratio in the UCC region increases from about 1.27 to 1.31 when the value of $L$ changes from 30 to 120 MeV, as a result of a larger $\Delta r_{\mathrm{np}}$ and thus a larger $\delta_{\rm spectator}$ for a larger $L$. Considering the deformation of $^{96}$Zr and $^{96}$Ru, the ratios in the UCC region from both $L$ are increased by about 0.01, so in this case the deformation effect is about $25\%$ the effect from $L$ but still controllable. The detecting efficiency for neutrons is about $100\%$, while that for protons is generally smaller since protons can be affected by the magnetic field due to its electric charge~\cite{Tarafdar:2014oua}. Experimentally, although one may try to measure the absolute multiplicity $N_p$ of protons and thus the $N_n/N_p$ ratio by correcting the detecting efficiency as in Fig.~\ref{fig:ratio}, it is easier and more accurate to measure the double ratio, i.e., the ratio of the $N_n/N_p$ ratio in the isobaric systems, so that the detecting efficiency for protons is effectively cancelled out. Similar method has been applied in low-energy collisions by different isotopes~\cite{Li:2005by,Famiano:2006rb}. The lower panel of Fig.~\ref{fig:ratio2} displays the double ratio of $N_n/N_p$ in isobaric collisions. Assuming that both $^{96}$Zr and $^{96}$Ru are spherical, the double ratio in the UCC region increases from about 1.6 to 1.7 when the value of $L$ changes from 30 to 120 MeV. Considering the deformation of $^{96}$Zr and $^{96}$Ru, the double ratios in the UCC region from both $L$ are decreased by less than 0.02. In this case the deformation effect is much smaller compared to that of $L$. It is also interesting to see that the effect of $L$ on the double ratio of $N_n/N_p$ in Fig.~\ref{fig:ratio2} (b) is about $6\%$ compared to the about $3\%$ effect of the $N_n$ ratio shown in Fig.~\ref{fig:ratio2} (a). Although the sensitivity of the double ratio of $N_n/N_p$ to $L$ is reduced compared to the single $N_n/N_p$ ratio, it is more easily measurable in experiments. The observables proposed in Fig.~\ref{fig:ratio2} can be generalized to any two colliding systems with similar mass number but different isospin asymmetries such as those along an isotopic chain, for which their sensitivity to $\Delta r_{\mathrm{np}}$ and $L$ is expected to increase with increasing difference in their isospin asymmetries.

\subsection{Excitation from electromagnetic field}

\begin{figure}[!h]
\includegraphics[width=0.8\linewidth]{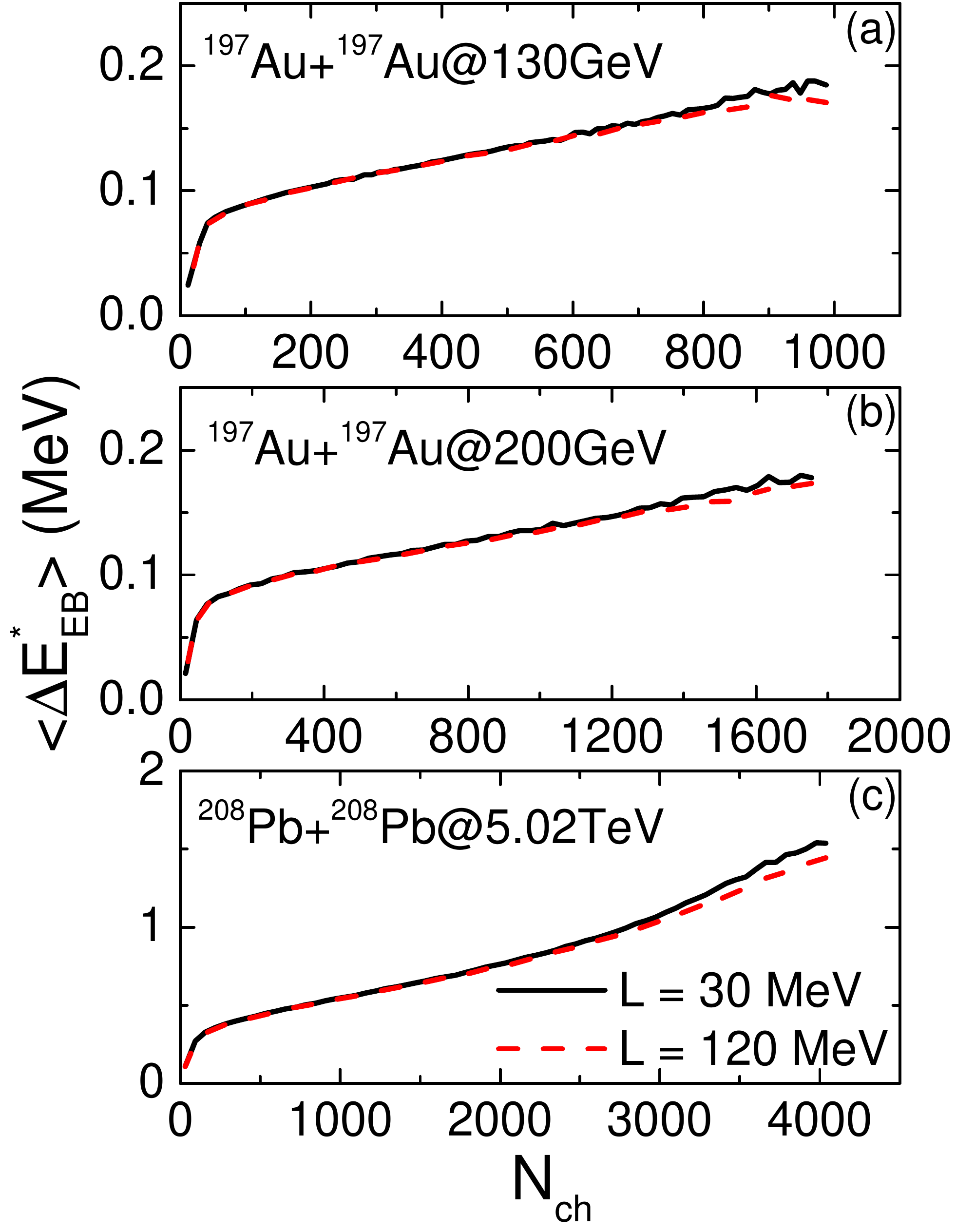}
\vspace*{-.3cm}
\caption{\label{fig:EBfield} Average energy excitation per spectator nucleon from the electromagnetic field as a function of $\nch$ in different collision systems and with different $L$.}
\end{figure}

At last, we discuss the possible effect from the electromagnetic field on the excitation energy of heavy clusters in spectator matter. As discussed in Sec.~\ref{sec:EBfield}, we have calculated the average increase of the excitation energy $\langle \Delta E_{\rm EB}^\star \rangle$ for each spectator nucleon from the increase of the kinetic energy of protons due to the retarded electromagnetic field, and its dependence on $\nch$ is shown in Fig.~\ref{fig:EBfield} for \auau\ at $\sqrtsnn=130$ and 200 GeV as well as for \pbpb\ at $\sqrtsnn=5.02$ TeV. One sees that $\langle \Delta E_{\rm EB}^\star \rangle$ increases with increasing $\nch$, and the difference from different $\Delta r_\mathrm{np}$ or $L$ is only visible at large $\nch$. The effects from the electromagnetic field for \auau\ at $\sqrtsnn=130$ and 200 GeV are much smaller than the assumed uncertainty of $\pm 1$ MeV per nucleon in calculating the excitation energy of heavy fragments. In contrast, the impact of electromagnetic field for \pbpb\ at $\sqrtsnn=5.02$ TeV is nearly comparable to the assumed uncertainty, and therefore should be investigated more quantitatively.

\begin{figure}[!h]
\includegraphics[width=0.8\linewidth]{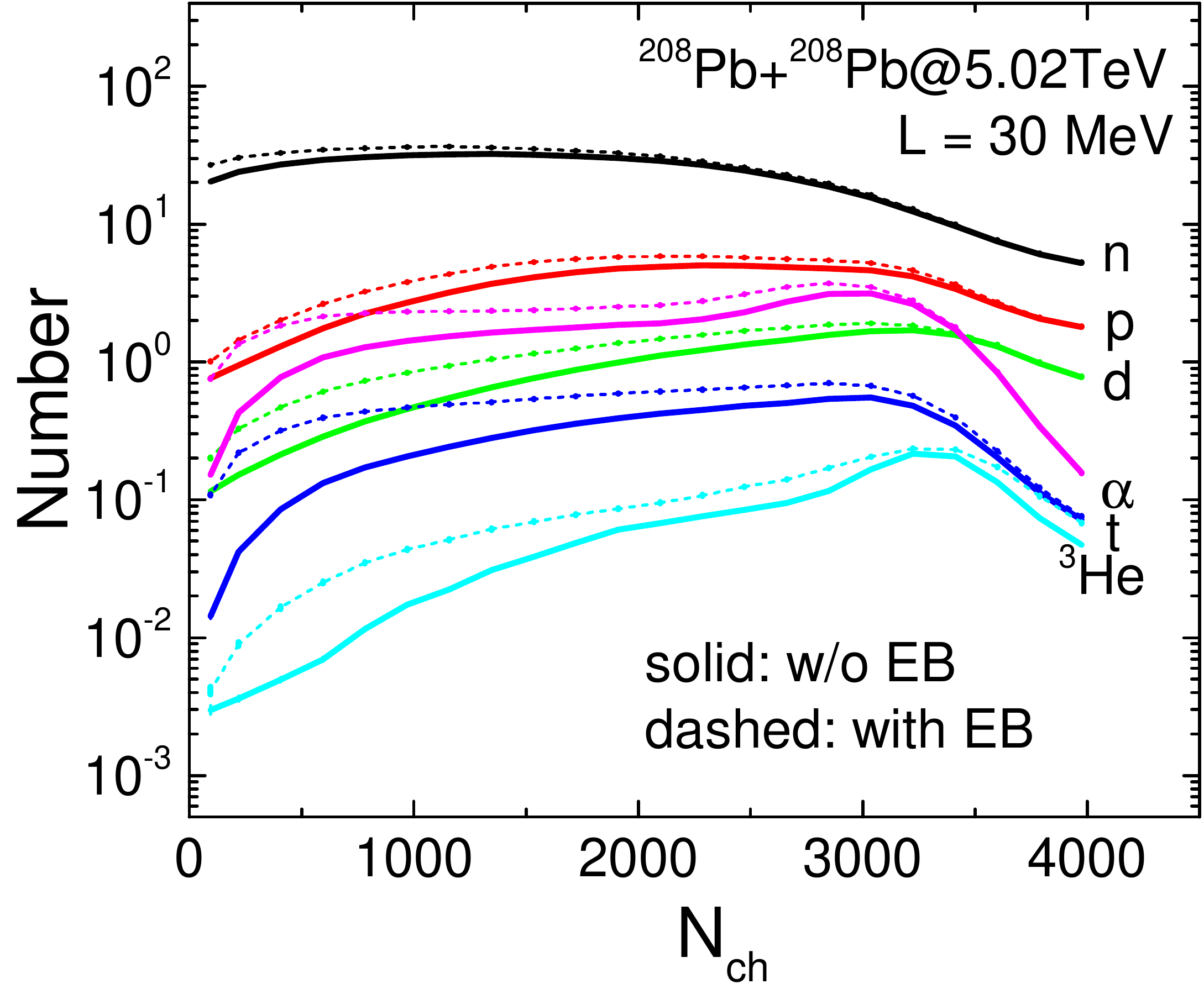}
\vspace*{-.3cm}
\caption{\label{fig:EBfield2} Numbers of free spectator nucleons and light clusters as a function of $\nch$ with and without the electromagnetic excitation in \pbpb\ collisions at $\sqrt{s_\mathrm{NN}}=5.02$ TeV for $L=30$ MeV.}
\end{figure}

The numbers of free spectator nucleons and light clusters with and without the electromagnetic excitation in \pbpb\ collisions at $\sqrt{s_\mathrm{NN}}=5.02$ TeV for $L=30$ MeV are compared in Fig.~\ref{fig:EBfield2}. One sees that the numbers of these particles are generally larger after considering the electromagnetic excitation. This is understandable since the excitation energy of heavy fragments becomes higher and they have chances to deexcite into more free nucleons or light clusters. Although the electromagnetic excitation is larger at larger $\nch$, its effect on the particle multiplicities is negligible there, since the deexcitation of heavy fragments is not important in the UCC region. We have also compared the results with the electromagnetic excitation for \auau\ collisions at $\sqrtsnn=130$ and 200 GeV as well as for \pbpb\ at $\sqrtsnn=5.02$ TeV in the panel (c), (d), and (e) of Fig.~\ref{fig:ratio}, respectively. As expected from the small excitation correction in \auau\ collisions at $\sqrtsnn=130$ and 200 GeV shown in Fig.~\ref{fig:EBfield}, the corresponding $N_n/N_p$ ratio is almost not affected. On the other hand, the $N_n/N_p$ ratio in \pbpb\ at $\sqrtsnn=5.02$ TeV is slightly smaller, as a result of more free nucleons as shown in Fig.~\ref{fig:EBfield2}.

\section{Summary and outlook}
\label{sec:summary}

We performed a detailed study of the production of free spectator nucleons and light nuclei in \zrzr, \ruru, \auau, and \pbpb\ collisions at RHIC and LHC energies. Besides the previously-studied ratio of free spectator neutrons in isobaric collisions~\cite{Liu:2022kvz}, we find that the yield ratio $N_n/N_p$  of free spectator neutrons to protons in a single collision system is positively correlated to the neutron-skin thickness $\Delta r_\mathrm{np}$ of colliding nuclei and the slope parameter $L$ of the nuclear symmetry energy. Therefore, a measurement of the $N_n/N_p$ in \pbpb\ collision at LHC energies could be used to probe the $\Delta r_\mathrm{np}$ in $^{208}$Pb and $L$ and compare with results from the recent PREXII experiment. Furthermore, the double ratio of $N_n/N_p$ in isobaric systems is insensitive to the detecting efficiency for protons and hence is a more reliable probe in future experiments. The above study is based on a Glauber model for nucleus-nucleus collisions, where proton and neutron distributions of the colliding nuclei are provided by the self-consistent spherical or deformed Skyrme-Hartree-Fock-Bogolyubov calculation, and final spectator particles are produced from direct production of the collisions, clusterization from the minimum spanning tree algorithm for heavy clusters and the Wigner function approach for light clusters, and deexcitation of heavy clusters by GEMINI in spectator matter. The proposed ratios in ultracentral collisions are most robust probes since there are few deexcitation processes there. The effect from the deformation of colliding nuclei is found to be secondary and controllable compared to the effect of $\Delta r_\mathrm{np}$ or $L$ on the observables. The electromagnetic excitation on the spectator matter is negligible at RHIC energies, but may have some influence at LHC energies.

In our theoretical study, the coalescence parameters for the formation of heavy clusters are tuned to describe the measured multiplicity of free spectator neutrons $N_n$ in ultracentral \auau\ collisions at $\sqrtsnn=130$ GeV~\cite{Liu:2022kvz}, but we underestimate the data in ultracentral \auau\ collisions at $\sqrtsnn=200$ GeV and \pbpb\ collisions at $\sqrtsnn=5.02$ TeV within the same approach. This difference could arise from excess neutral particles besides spectator neutrons that may fall into the acceptance of the zero-degree calorimeters. To make good use of the probes proposed in the present study, this possible background contribution needs further investigations.

\begin{acknowledgments}
JX is supported by the National Natural Science Foundation of China under Grant No. 11922514. JJ and CZ are supported by the US Department of Energy under Contract No. DEFG0287ER40331. GXP and LML are supported by the National Natural Science Foundation of China under Grant Nos. 11875052, 11575190, and 11135011.
\end{acknowledgments}

\bibliography{isobar-long}
\end{document}